\documentclass[letterpaper,twocolumn,10pt]{article}
\usepackage[svgnames]{xcolor}
\usepackage{usenix-2020-09}

\usepackage{cite}
\usepackage{amsmath,amssymb,amsfonts}
\usepackage{algorithmic}
\usepackage{multirow}
\usepackage{graphicx}
\usepackage{textcomp}
\usepackage{xcolor}
\usepackage{booktabs}
\usepackage{enumitem}
\def\BibTeX{{\rm B\kern-.05em{\sc i\kern-.025em b}\kern-.08em
    T\kern-.1667em\lower.7ex\hbox{E}\kern-.125emX}}
\usepackage{xspace} 
\usepackage{pifont}
\usepackage[many]{tcolorbox}

\usepackage{tikz}
\usetikzlibrary{positioning,shapes,arrows.meta,fit}
\usepackage{dblfloatfix}    
\usepackage{caption}
\usepackage{subcaption}

\usepackage{colortbl}
\usepackage[noblocks]{authblk}

\newcommand{\gptthree}{\textsc{GPT-3.5}\xspace}
\newcommand{\gptfour}{\textsc{GPT-4}\xspace}

\newtheorem{myDef}{Definition}

\newcommand*{\pttab}{{{PTT}}\xspace}

\newcommand{\tool}{\textsc{PentestGPT}}
\newcommand{\toolgptthree}{\textsc{PentestGPT-GPT-3.5}}
\newcommand{\toolgptfour}{\textsc{PentestGPT-GPT-4}}
\newcommand{\toolnoreasoning}{\textsc{PentestGPT-no-Reasoning}}
\newcommand{\toolnogeneration}{\textsc{PentestGPT-no-Generation}}
\newcommand{\toolnoparsing}{\textsc{PentestGPT-no-Parsing}}

\newcommand{\framework}{\textsc{Malism}}
\newcommand{\ef}{\textsc{ExploitFlow}}
\newcommand{\pentestperf}{\textsc{PentestPerf}}

\newcommand{\edit}[1]{{#1}}

\definecolor{darkred}{HTML}{860000}
\definecolor{darkteal}{HTML}{005959}
\definecolor{darkpurple}{HTML}{590059}
\definecolor{darkgrey}{HTML}{434343}

\newtcolorbox{mybox}[2][]{text width=0.95\linewidth,fontupper=\normalsize,
fonttitle=\bfseries\sffamily\scriptsize, colbacktitle=darkgrey,enhanced,
attach boxed title to top left={yshift=-2mm,xshift=3mm},
boxed title style={sharp corners},top=4pt,bottom=2pt,left=2pt,right=2pt,
  title=#2,colback=white}

\newtcolorbox{gptbox}[2][]{text width=0.95\linewidth,fontupper=\normalsize,
fonttitle=\bfseries\sffamily\scriptsize, colbacktitle=darkgrey,enhanced,
attach boxed title to top right={yshift=-2mm,xshift=-3mm},
boxed title style={sharp corners},top=4pt,bottom=2pt,left=2pt,right=2pt,
  title=#2,colback=white}

\begin{document}

\title{\tool{}: Evaluating and Harnessing Large Language Models for Automated Penetration Testing}

\newcommand{\mkntu}[0]{{{$^1$}}}
\newcommand{\mkal}[0]{{{$^2$}}}
\newcommand{\mkaau}[0]{{{$^3$}}}
\newcommand{\mkitr}[0]{{{$^4$}}}
\newcommand{\mkunsw}[0]{{{$^5$}}}
\newcommand{\mkjku}[0]{{{$^6$}}}

\author{
    {\rm Gelei Deng}\mkntu \rm ,
    {\rm Yi Liu}\mkntu \rm ,
    {\rm Víctor Mayoral-Vilches}\mkal \rm \mkaau \rm ,
    {\rm Peng Liu}\mkitr \rm ,
    {\rm Yuekang Li}\mkunsw \rm ,
    {\rm Yuan Xu}\mkntu \rm ,\linebreak
    {\rm Tianwei Zhang}\mkntu \rm , 
    {\rm Yang Liu}\mkntu \rm , 
    {\rm Martin Pinzger}\mkaau \rm , 
    {\rm Stefan Rass}\mkjku \rm \\
    \mkntu {Nanyang Technological University},
    \mkal {Alias Robotics},
    \mkaau {Alpen-Adria-Universität Klagenfurt},\linebreak
    \mkitr {Institute for Infocomm Research (\(I^2R\)), A*STAR, Singapore},
    \mkunsw {University of New South Wales},
    \mkjku {Johannes Kepler University Linz}\\
    \medskip

}

\maketitle
\thispagestyle{plain}
\pagestyle{plain}

\begin{abstract}

Penetration testing, a crucial industrial practice for ensuring system security, has traditionally resisted automation due to the extensive expertise required by human professionals. 
Large Language Models (LLMs) have shown significant advancements in various domains, and their emergent abilities suggest their potential to revolutionize industries. 
In this work, we establish a comprehensive benchmark using real-world penetration testing targets and further use it to explore the capabilities of LLMs in this domain.
Our findings reveal that while LLMs demonstrate proficiency in specific sub-tasks within the penetration testing process, such as using testing tools, interpreting outputs, and proposing subsequent actions, they also encounter difficulties maintaining a whole context of the overall testing scenario.

Based on these insights, we introduce \tool{}, an LLM-empowered automated penetration testing framework that leverages the abundant domain knowledge inherent in LLMs. 
\tool{} is meticulously designed with three self-interacting modules, each addressing individual sub-tasks of penetration testing, to mitigate the challenges related to context loss. 
Our evaluation shows that \tool{} not only outperforms LLMs with a task-completion increase of 228.6\% compared to the \gptthree model among the benchmark targets, but also \edit{proves effective in tackling real-world penetration testing targets and CTF challenges}. 
\edit{Having been open-sourced on GitHub, \tool{} has garnered over 6,200 stars in 9 months and fostered active community engagement, attesting to its value and impact in both the academic and industrial spheres.}

\end{abstract}

\section{Introduction}



Securing a system presents a formidable challenge. Offensive security methods like penetration testing (pen-testing) and red teaming are now essential in the security lifecycle. As explained by Applebaum~\cite{applebaum2017analysis}, these approaches involve security teams attempting breaches to reveal vulnerabilities, providing advantages over traditional defenses, which rely on incomplete system knowledge and modeling. This study, guided by the principle \emph{``the best defense is a good offense''}, focuses on offensive strategies, specifically penetration testing.

Penetration testing is a proactive offensive technique for identifying, assessing, and mitigating security vulnerabilities~\cite{arkin2005software}.
It involves targeted attacks to confirm flaws, yielding a comprehensive inventory of vulnerabilities with actionable recommendations. 
This widely-used practice empowers organizations to detect and neutralize network and system vulnerabilities before malicious exploitation. 
However, it typically relies on manual effort and specialized knowledge~\cite{dengnautilus}, resulting in a labor-intensive process, creating a gap in meeting the growing demand for efficient security evaluations.


Large Language Models (LLMs) have demonstrated profound capabilities, showcasing intricate comprehension of human-like text and achieving remarkable results across a multitude of tasks~\cite{zhao2023survey, liu2023summary}. An outstanding characteristic of LLMs is their emergent abilities~\cite{wei2022emergent}, cultivated during training, which empower them to undertake intricate tasks such as reasoning, summarization, and domain-specific problem-solving without task-specific fine-tuning. This versatility posits LLMs as potential game-changers in various fields, notably cybersecurity.
Although recent  works~\cite{mayoralvilches2023exploitflow, zhang2023does,he2024large} posit the potential of LLMs to reshape cybersecurity practices, including the context of penetration testing, there is an absence of a systematic, quantitative assessment of their aptitude in this regard. Consequently, an imperative question presents: To what extend can LLMs automate penetration testing?


Motivated by this question, we set out to explore the capability boundary of LLMs on real-world penetration testing tasks. Unfortunately, the current benchmarks for penetration testing~\cite{antunes2010benchmarking, xiong2010model} are not comprehensive and fail to assess progressive accomplishments fairly during the process. To address this limitation, we construct a robust benchmark that includes test machines from HackTheBox~\cite{htb} and VulnHub~\cite{vulnhub}—two leading platforms for penetration testing challenges. Comprising 13 targets with 182 sub-tasks, our benchmark encompasses all vulnerabilities appearing in OWASP's top 10 vulnerability list~\cite{owasp} and 18 Common Weakness Enumeration (CWE) items~\cite{cwe}. The benchmark offers a more detailed evaluation of the tester's performance by monitoring the completion status for each sub-task.

With this benchmark, we perform an exploratory study using GPT-3.5~\cite{gpt3.5}, GPT-4~\cite{gpt4}, and Bard~\cite{bard} as representative LLMs. 
Our test strategy is interactive and iterative. We craft tailored prompts to guide the LLMs through penetration testing. 
Each LLM, presented with prompts and target machine information, generates step-by-step penetration testing operations. We then execute the suggested operations in a controlled environment, document the results, and feed them back to the LLM to inform and refine its next steps. This cycle (prompting, executing, and feedback) is repeated until the LLM completes the entire penetration testing process autonomously. To evaluate LLMs, we compare their results against baseline solutions from official walkthroughs and certified penetration testers. By analyzing similarities and differences in their problem-solving approaches, we aim to better understand LLMs' capabilities in penetration testing and how their strategies differ from human experts.

Our investigation yields intriguing insights into the capabilities and limitations of LLMs in penetration testing. We discover that LLMs demonstrate proficiency in managing specific sub-tasks within the testing process, such as utilizing testing tools, interpreting their outputs, and suggesting subsequent actions. Compared to human experts, LLMs are especially adept at executing complex commands and options with testing tools, while models like \gptfour{} excel in comprehending source code and pinpointing vulnerabilities. Furthermore, LLMs can craft appropriate test commands and accurately describe graphical user-interface operations needed for specific tasks. Leveraging their vast knowledge base, they can design inventive testing procedures to unveil potential vulnerabilities in real-world systems and CTF challenges. However, we also note that LLMs have difficulty in maintaining a coherent grasp of the overarching testing scenario, a vital aspect for attaining the testing goal. As the dialogue advances, they may lose sight of earlier discoveries and struggle to apply their reasoning consistently toward the final objective. Additionally, LLMs overemphasize recent tasks in the conversation history, regardless of their vulnerability status. As a result, they tend to neglect other potential attack surfaces exposed in prior tests and fail to complete the penetration testing task.

Building on our insights into LLMs' capabilities in penetration testing, we present \tool{}\footnote{\tool{} is King Arthur's legendary sword, known for its exceptional cutting power and the ability to pierce armor.}, an interactive system designed to enhance the application of LLMs in this domain. Drawing inspiration from the collaborative dynamics commonly observed in real-world human penetration testing teams, \tool{} is particularly tailored to manage large and intricate projects. It features a tripartite architecture comprising \textit{Reasoning, Generation, and Parsing Modules}, each reflecting specific roles within penetration testing teams.
The Reasoning Module emulates the function of a lead tester, focusing on maintaining a high-level overview of the penetration testing status. We introduce a novel representation, the Pentesting Task Tree (PTT), based on the cybersecurity attack tree~\cite{attacktree}. This structure encodes the testing process's ongoing status and steers subsequent actions. Uniquely, this representation can be translated into natural language and interpreted by the LLM, thereby comprehended by the Generation Module and directing the testing procedure.
The Generation Module, mirroring a junior tester's role, is responsible for constructing detailed procedures for specific sub-tasks. Translating these into exact testing operations augments the generation process's accuracy. Meanwhile, the Parsing Module deals with diverse text data encountered during penetration testing, such as tool outputs, source codes, and HTTP web pages. It condenses and emphasizes these texts, extracting essential information. Collectively, these modules function as an integrated system. \tool{} completes complex penetration testing tasks by bridging high-level strategies with precise execution and intelligent data interpretation, thereby maintaining a coherent and effective testing process.

\edit{We assessed \tool{} across diverse testing scenarios to validate its effectiveness and breadth. In our custom benchmarks, \tool{} significantly outperformed direct applications of \gptthree{} and \gptfour{}, showing increases in sub-task completion rates of 228.6\% and 58.6\%, respectively. Furthermore, when applied to real-world challenges such as the HackTheBox active machine penetration tests~\cite{htbactive} and picoMini~\cite{picoCTF} CTF competition, \tool{} demonstrated its practical utility. It successfully resolved 4 out of 10 penetration testing challenges, incurring a total cost of 131.5 US Dollars for the OpenAI API usage. In the CTF competition, \tool{} achieved a score of 1500 out of a possible 4200, placing 24th among 248 participating teams.}
This evaluation underscores \tool{}'s practical value in enhancing penetration testing tasks' efficiency and precision. The solution has been made publicly available on GitHub\footnote{The project is at: \url{https://github.com/GreyDGL/PentestGPT}.}, receiving widespread acclaim with over \edit{6,200} stars to the date of writing, active community engagement, and ongoing collaboration with multiple industrial partners.

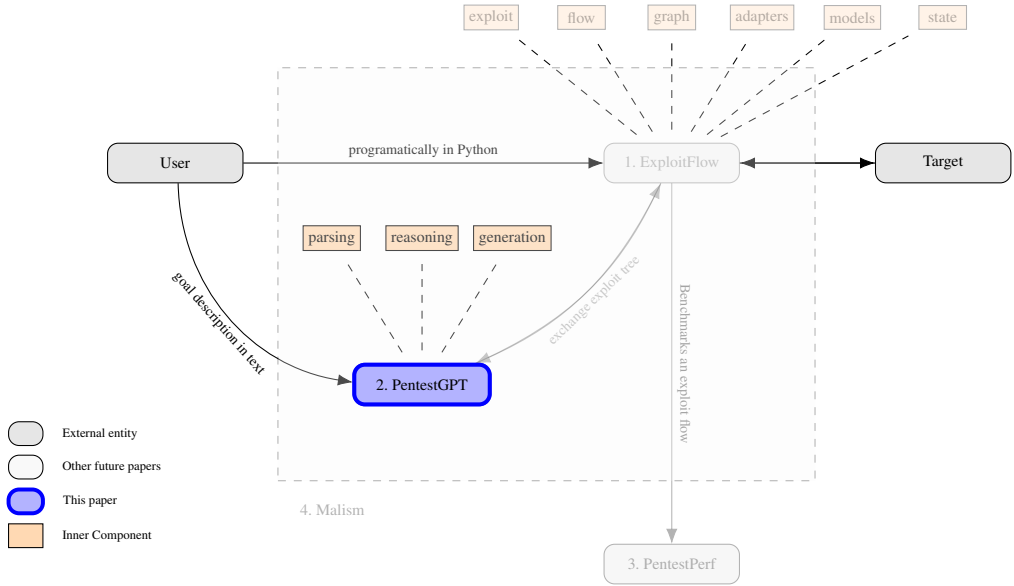
\begin{figure*}[t]
    \centering 
    \begin{tikzpicture}[node distance=2cm, auto, scale=0.6, transform shape]
    \tikzstyle{box} = [rectangle, draw, fill=gray!20, text centered, rounded corners, minimum height=2.5em, minimum width=3cm]
    \tikzstyle{emphasize} = [box, fill=blue!30, draw=blue, line width=1.5pt]
    \tikzstyle{line} = [draw, -{Latex[scale=1.2]}]
    \tikzstyle{deemphasize} = [opacity=0.3]
    \tikzstyle{innerComp} = [rectangle, draw, fill=orange!30, text centered, minimum height=1.5em, minimum width=3em]

    \node[box] (U) {User};
    \node[box, deemphasize] (EF) at (11cm,0) {1. ExploitFlow}; 
    \node[box, emphasize, below left=4cm and 2.5cm of EF] (PentestGPT) {2. PentestGPT};
    \node[box, deemphasize, below=8cm of EF] (PentestPerf) {3. PentestPerf};
    \node[box, right=3cm of EF] (T) {Target};

    \node[innerComp, deemphasize, above=2.5cm of EF, xshift=-4cm] (exploit) {exploit};
    \node[innerComp, deemphasize, right of=exploit] (flow) {flow};
    \node[innerComp, deemphasize, right of=flow] (graph) {graph};
    \node[innerComp, deemphasize, right of=graph] (adapters) {adapters};
    \node[innerComp, deemphasize, right of=adapters] (models) {models};
    \node[innerComp, deemphasize, right of=models] (state) {state};

    \node[innerComp, above=2.5cm of PentestGPT, xshift=-2cm] (parsing) {parsing};
    \node[innerComp, right of=parsing] (reasoning) {reasoning};
    \node[innerComp, right of=reasoning] (generation) {generation};

    \draw[dashed, shorten <=5pt, shorten >=5pt] (EF) -- (exploit);
    \draw[dashed, shorten <=5pt, shorten >=5pt] (EF) -- (flow);
    \draw[dashed, shorten <=5pt, shorten >=5pt] (EF) -- (graph);
    \draw[dashed, shorten <=5pt, shorten >=5pt] (EF) -- (adapters);
    \draw[dashed, shorten <=5pt, shorten >=5pt] (EF) -- (models);
    \draw[dashed, shorten <=5pt, shorten >=5pt] (EF) -- (state);

    \draw[dashed, shorten <=5pt, shorten >=5pt] (PentestGPT) -- (parsing);
    \draw[dashed, shorten <=5pt, shorten >=5pt] (PentestGPT) -- (reasoning);
    \draw[dashed, shorten <=5pt, shorten >=5pt] (PentestGPT) -- (generation);

    \path[line] (U) -- node[font=\small, midway, sloped, above] {programatically in Python} (EF); 
    \path[line] (U) edge[bend right=40] node[font=\small, sloped, below] {goal description in text} (PentestGPT);
    \path[line, deemphasize] (PentestGPT) edge[bend right=20] node[font=\small, sloped, below] {exchange exploit tree} (EF);
    \path[line, deemphasize] (EF) edge[bend left=20] node {} (PentestGPT);
    \path[line, deemphasize] (EF) -- node[font=\small, sloped, above, opacity=0.5] {Benchmarks an exploit flow} (PentestPerf);
    \path[line] (EF) -- (T);
    \path[line] (T) -- (EF);

    \node[rectangle, fill=gray!5, draw, dashed, deemphasize, fit=(EF) (PentestGPT), inner sep=1cm] (group) {};
    \node[anchor=north west, deemphasize, inner sep=0.5cm] at (group.south west) {4. Malism};

    \node[box, emphasize, below left=4cm and 2.5cm of EF] (PentestGPT) {2. PentestGPT};

    \begin{scope}[node distance=0.3cm, font=\footnotesize]
        \coordinate (legendOrigin) at (-4,-6); 
        \node[box, right=0.3cm of legendOrigin, minimum height=1.5em, minimum width=1.5em, text width=1.5em] (legendBox) {}; 
        \node[right=0.3cm of legendBox] {External entity};

        \node[box, below=0.2cm of legendBox, fill=gray!5, minimum height=1.5em, minimum width=1.5em, text width=1.5em] (otherpapers) {}; 
        \node[right=0.3cm of otherpapers] {Other future papers};

        \node[emphasize, below=0.2cm of otherpapers, minimum height=1.5em, minimum width=1.5em, text width=1.5em] (legendEmphasize) {};
        \node[right=0.3cm of legendEmphasize] {This paper};

        \node[innerComp, below=0.2cm of legendEmphasize, minimum height=1.5em, minimum width=1.5em, text width=1.5em] (legendInner) {};
        \node[right=0.3cm of legendInner] {Inner Component};
    \end{scope}
\end{tikzpicture}
    \caption{Architecture of our framework to develop a fully automated penetration testing tools, \framework{}. Figure depicts the various interaction flows that an arbitrary \texttt{User} could follow using \framework{} to pentest a given \texttt{Target}. \textbf{1.} Corresponds with \ef{}, a modular library to produce security exploitation routes (\emph{exploit flows}) that caputures the state of the system being tested in a flow after every discrete action. \textbf{2. (this paper)} Corresponds with \tool{}, a testing tool that leverages the power of LLMs to produce testing guidance (heuristics) for every given discrete state. \textbf{3.} \pentestperf~is a comprehensive penetration testing benchmark to evaluate the performances of penetration testers and automated tools across a wide array of testing targets. \textbf{4.} captures \framework{}, our framework to develop fully automated penetration testing tools which we name \emph{cybersecurity cognitive engines}.
    }
    \label{fig:my_figure_label} 
\end{figure*}

As a long term research goal, we aim to contribute to unlocking the potential of modern machine learning approaches and develop a fully automated penetration testing framework that helps produce cybersecurity cognitive engines. Our overall architecture is depicted in Figure~\ref{fig:my_figure_label}, showing our current work and future planned contributions.
Our proposed framework, \framework{}, is designed to enable a user without in-depth security domain knowledge to produce its cybersecurity cognitive engine that helps conduct penetration testing over an extensive range of targets. This framework comprises three primary components:
\begin{enumerate}

\item \ef{} \cite{mayoral2023exploitflow}: A modular library to produce cyber security exploitation routes (\emph{exploit flows}). \ef{} aims to combine and compose exploits from different sources and frameworks, capturing the state of the system being tested in a flow after every discrete action, which allows learning attack trees that affect a given system. \ef{}'s main motivation is to facilitate and empower Game Theory and Artificial Intelligence (AI) research in cyber security. It uniquely represents the exploitation process that encodes every facet within it. Its representation can be effectively integrated with various penetration testing tools and scripts, such as Metasploit~\cite{metasploit} to perform end-to-end penetration testing. Such representation can be further visualized to guide the human experts to reproduce the testing process. 

\item \tool{} (\textbf{this paper}): An automated penetration testing system that leverages the power of LLMs to produce testing guidance and intuition at every given discrete state. It functions as the core component of the \framework{} framework, guiding the LLMs to utilize their domain knowledge in real-world testing scenarios efficiently.

\item \pentestperf: A comprehensive penetration testing benchmark developed to evaluate the performances of penetration testers and automated tools across a wide array of testing targets. It offers a fair and robust platform for performance comparison.
\end{enumerate}

The harmonious integration of these three components forms an automated, self-evolving penetration testing framework capable of executing penetration tests over various targets, \framework{}. This framework to develop fully automated penetration testing tools, which we name \emph{cybersecurity cognitive engines}, aims to revolutionize the field of penetration testing by significantly reducing the need for domain expertise and enabling more comprehensive and reliable testing.

In summary, we make the following contributions:

\begin{itemize}[leftmargin=*]

\item \textbf{Development of a Comprehensive Penetration Testing Benchmark.} We craft a robust and representative penetration testing benchmark, encompassing a multitude of test machines from leading platforms such as HackTheBox and VulnHub. This benchmark includes 182 sub-tasks covering OWASP's top 10 vulnerabilities, offering fair and comprehensive evaluation of penetration testing.
To the best of our knowledge, this is the first benchmark in the field that can provide progressive accomplishments assessments and comparisons.

\item \textbf{Comprehensive Evaluation of LLMs for Penetration Testing Tasks.} By employing models like \gptthree{}, \gptfour{}, and Bard, our exploratory study rigorously investigates the strengths and limitations of LLMs in penetration testing. 
To the best of our knowledge, this is the first systematic and quantitative study for the capability of LLMs in performing automated penetration testing.
The insights gleaned from this study shed valuable light on the capabilities and challenges faced by LLMs, enriching our understanding of their applicability in this specialized domain.

\item \textbf{Development of an Innovative LLM-powered Penetration Testing System.} We engineer \tool{}, a novel interactive system that leverages the strengths of LLMs to carry out penetration testing tasks automatically. Drawing inspiration from real-world human penetration testing teams, \tool{} integrates a tripartite design that mirrors the collaborative dynamics between senior and junior testers. This architecture optimizes LLMs' usage, significantly enhancing the efficiency and effectiveness of automated penetration testing.
We have open-sourced \tool{} and it has received over 6,500 stars on GitHub, active community contributions, and industry partners including AWS, Huawei, and ByteDance to collaborate.

\end{itemize}

\section{Background \& Related Work}\label{sec:background}
\subsection{Penetration Testing}
Penetration testing, or ``pentesting'', is a critical practice to enhance organizational systems' security. In a typical penetration test, security professionals, known as penetration testers, analyze the target system, often leveraging automated tools. The standard process is divided into five key phases~\cite{weidman2014penetration}: Reconnaissance, Scanning, Vulnerability Assessment, Exploitation, and Post Exploitation (including reporting). These phases enable testers to understand the target system, identify vulnerabilities, and exploit them to gain access.

Despite significant advancements~\cite{abu2018automated, schwartz2019autonomous, xiong2010model}, a fully automated penetration testing system remains out of reach. This gap results from the need for deep vulnerability understanding and a strategic action plan. Typically, testers combine depth-first and breadth-first search techniques~\cite{weidman2014penetration}. They first grasp the target environment's scope, then drill down into specific vulnerabilities. This method ensures comprehensive analysis, leaning on expertise and experience. The multitude of specialized tools further complicate the automation. Thus, even with artificial intelligence, achieving a seamless automated penetration testing solution is a daunting task.

\subsection{Large Language Models}
Large Language Models (LLMs), including OpenAI's \gptthree{} and \gptfour{}, are prominent tools with applications extending to various cybersecurity-related fields, such as code analysis~\cite{pearce2022asleep} and vulnerability repairment~\cite{pearce2023examining}. These models are equipped with wide-ranging general knowledge and the capacity for elementary reasoning. They can comprehend, infer, and produce text resembling human communication, aided by a training corpus encompassing diverse domains like computer science and cybersecurity. Their ability to interpret context and recognize patterns enables them to adapt knowledge to new scenarios. This adaptability, coupled with their proficiency in interacting with systems in a human-like way, positions them as valuable assets in enhancing penetration testing processes. Despite inherent limitations, LLMs offer distinct attributes that can substantially aid in the automation and improvement of penetration testing tasks. The realization of this potential, however, requires the creation and application of a specialized and rigorous benchmark.

\section{Penetration Testing Benchmark}\label{sec:benchmark}
\subsection{Motivation}\label{sec:benchmark:motivation}
The comprehensive evaluation of LLMs in penetration testing necessitates a robust and representative benchmark. Existing benchmarks in this domain~\cite{antunes2010benchmarking,xiong2010model} have several limitations. 
\edit{First, they are often restricted in scope, focusing on a narrow range of potential vulnerabilities, and thus fail to capture the complexity and diversity of real-world cyber threats. For instance, the OWASP \textit{juiceshop} project~\cite{owasp-juiceshop} is the most widely adopted benchmark for web vulnerability evaluation. However, it does not include privilege escalation vulnerabilities, which is an essential aspect of penetration testing.}
Second, existing benchmarks may not recognize the cumulative value of progress through the different stages of penetration testing, as they tend to evaluate only the final exploitation success. This approach overlooks the nuanced value each step contributes to the overall process, resulting in metrics that might not accurately represent actual performance in real-world scenarios.

To address these concerns, we propose the construction of a comprehensive penetration testing benchmark that meets the following criteria:

\noindent\textbf{Task Variety.} The benchmark must encompass diverse tasks, reflecting various operating systems and emulating the diversity of scenarios encountered in real-world penetration testing.

\noindent\textbf{Challenge Levels.} To ensure broad applicability, the benchmark must include tasks of varying difficulty levels suitable for challenging novice and expert testers.

\noindent\textbf{Progress Tracking.} Beyond mere success or failure metrics, the benchmark must facilitate tracking of incremental progress, thereby recognizing and scoring the value added at each stage of the penetration testing process.

\subsection{Benchmark Design}\label{sec:benchmark:design}
Following the criteria outlined previously, we develop a comprehensive benchmark that closely reflects real-world penetration testing tasks. The design process progresses through several stages.

\noindent\textbf{Task Selection.} We begin by selecting tasks from HackTheBox~\cite{htb} and VulnHub~\cite{vulnhub}, \edit{two leading penetration testing training platforms. Our selection criteria are designed to ensure that our benchmark accurately reflects the challenges encountered in practical penetration testing environments. We meticulously review the latest machines available on both platforms, aiming to identify and select a subset that comprehensively covers all vulnerabilities listed in the OWASP~\cite{owasp} Top 10 Project. Additionally, we choose machines that represent a mix of difficulties, classified according to traditional standards in the penetration testing domain into \textit{easy}, \textit{medium}, and \textit{hard} categories. This process guarantees that our benchmark spans the full spectrum of vulnerabilities and difficulties. } Note that our benchmark does not include benign targets to assess false positives. In penetration testing, benign targets are sometimes explored. Our main objective remains identifying true vulnerabilities.

\noindent\textbf{Task Decomposition.} We further parse the testing process of each target into a series of sub-tasks, following the standard solution commonly referred to as the ``walkthrough'' in penetration testing. 
Each sub-task corresponds to a unique step in the overall process. We decompose sub-tasks following NIST 800-115~\cite{nist2012nist}, the Technical Guide to Security Testing. Each sub-task is one step declared in the Guide (e.g., network discovery, password cracking), or an operation that exploits a unique vulnerability categorised in the Common Weakness Enumeration (CWE)~\cite{cwe} (e.g., exploiting SQL injection - CWE-89~\cite{cwe-sql}).
In the end, we formulate an exhaustive list of sub-tasks for every benchmark target. We provide the complete list of the decomposed sub-tasks in Appendix Table~\ref{tab:subtask-cat}.

\noindent\textbf{Benchmark Validation.} \edit{The final stage of our benchmark development involves rigorous validation, which ensures the reproducibility of these benchmark machines. To do this, three certified penetration testers independently attempt the penetration testing targets and write their walkthrough. We then adjust our task decomposition accordingly because some targets may have multiple valid solutions.}

\edit{Ultimately, we have compiled a benchmark that effectively encompasses all types of vulnerabilities listed in the OWASP~\cite{owasp} Top 10 Project. It comprises 13 penetration testing targets, each at varying levels of difficulty. These targets are broken down into 182 sub-tasks across 26 categories, covering 18 distinct CWE items. This number of targets is deemed sufficient to represent a broad spectrum of vulnerabilities, difficulty levels, and varieties essential for comprehensive penetration testing training. Detailed information about the included categories can be found in the Appendix Section~\ref{tab:subtask-cat}. To foster community development, we have made this benchmark publicly available online at our anonymous project website~\cite{anonymous-project}.}
\section{Exploratory Study}\label{sec:exploratory-study}

We conduct an exploratory study to assess the capabilities of LLMs in penetration testing, \edit{with the  primary objective of determining how well LLMs can adapt to the real-world complexities and challenges in this task.} Specifically, we aim to address the following two research questions:

\noindent\textbf{RQ1 (Capability):} To what extent can LLMs perform penetration testing tasks? 

\noindent\textbf{RQ2 (Comparative Analysis):} How do the problem-solving strategies of human penetration testers and LLMs differ?

We utilize the benchmark described in Section~\ref{sec:benchmark} to evaluate the performance of LLMs on penetration testing tasks. In the following, we first delineate our testing strategy for this study. Subsequently, we present the testing results and an analytical discussion to address the above research questions.

\subsection{Testing Strategy}\label{sec:exploratory-study:testing-strategy}

\edit{LLMs are text-based and cannot independently perform penetration testing operations. To address this, we develop a human-in-the-loop testing strategy, serving as an intermediary method to accurately assess LLMs' capabilities. This strategy features an interactive loop where a human expert executes the LLM's penetration testing directives. Importantly, the human expert functions purely as an executor, strictly following the LLM's instructions without adding any expert insights or making independent decisions.}

\begin{figure}[t]
	\centering
	\includegraphics[width=\linewidth]{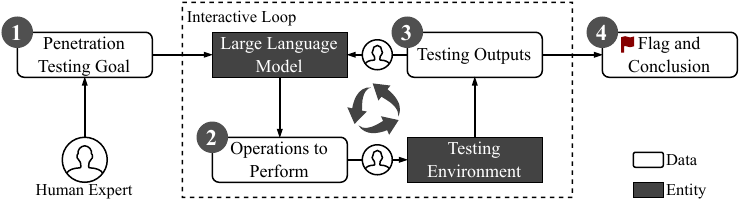}
 \caption{Overview of strategy to use LLMs for penetration testing.}
	\label{fig:testing-strategy}
\end{figure}

\edit{Figure~\ref{fig:testing-strategy} decipits the testing strategy with the following steps}: \ding{182}  We initiate the looped testing procedure by presenting the target specifics to the LLM, seeking its guidance on potential penetration testing steps.
\ding{183} The human expert strictly follows the LLM's recommendations and conducts the suggested actions in the penetration testing environment.
\ding{184} Outcomes of the testing actions are collected and summarized: direct text outputs such as terminal outputs or source code are documented; non-textual results, such as graphical representations, are translated by the human expert into succinct textual summaries. The data is then fed back to the LLM, setting the stage for its subsequent recommendations.
\ding{185} This iterative process persists either until a conclusive solution is identified or an deadlock is reached. We then compile a record of the testing procedures, encompassing successful sub-tasks, ineffective actions, and any reasons for failure, if applicable. 
For a more tangible grasp of this strategy, we offer illustrative examples of prompts and corresponding outputs from \gptfour{} related to one of our benchmark targets in the Appendix Section~\ref{appendix:example}.

\edit{To ensure the evaluation's fairness and accuracy, we employ several strategies. First, we involve expert-level penetration testers\footnote{We selected Offensive Security Certified Professionals (OSCP) testers.} as the human testers. With their deep pentesting knowledge, these testers can precisely comprehend and execute LLM-generated operations, thus accurately assessing LLMs' true capabilities.}
\edit{Second, we instruct the penetration testers to strictly execute the commands given by the LLMs, without altering any content or information, even upon identifying clear errors. They are also instructed to faithfully report the testing results back to the LLM without any additional commentary.}
\edit{Third, for managing UI-based operations and graphical results, we have adopted specific measures. Initially, we instruct the LLMs to minimize the use of GUI-based tools. For indispensable tools that cannot be avoided (e.g., BurpSuite), we propose a result-oriented approach: upon receiving a GUI operation instruction, the testers first execute the operation based on their expert knowledge. Subsequently, they are required to provide detailed, step-by-step textual descriptions of their actions and the observed responses at each step, which are then communicated back to the LLM. Should the LLM express any objections or comments concerning a particular step, the operation is to be repeated. This protocol ensures the integrity of the feedback loop, guaranteeing that the LLM obtains a comprehensive understanding of the testing results.}

\begin{table*}[th]
\caption{\edit{Overall performance of LLMs on Penetration Testing Benchmark.}}
\resizebox{\linewidth}{!}{
\centering
\begin{tabular}{c|cc|cc|cc|cc} 
\toprule
\multicolumn{1}{c}{ }        & \multicolumn{2}{c}{\textbf{Easy}}                                           & \multicolumn{2}{c}{\textbf{Medium}}                                           & \multicolumn{2}{c}{\textbf{Hard}}                                       & \multicolumn{2}{c}{\textbf{Average}}                                           \\ \midrule
Tools                       & Overall (7)                      & Sub-task (77)                   & Overall (4)                      & Sub-task (71)                     & Overall (2)                   & Sub-task (34)                  & \multicolumn{1}{c}{Overall (13)} & \multicolumn{1}{c}{Sub-task (182)} \\ \midrule
\gptthree{}                      & 1 (14.29\%)                       & 24 (31.17\%)                     & 0 (0.00\%)                        & 13 (18.31\%)                       & 0 (0.00\%)                     & 5 (14.71\%)                     & 1 (7.69\%)                        & 42 (23.07\%)                        \\
\gptfour{}                        & 4 (57.14\%)                       & \edit{55 (71.43\%)}                     & 1 (25.00\%)                       & \edit{30 (42.25\%)}                       & 0 (0.00\%)                     & \edit{10 (29.41\%)}                     & 5 (38.46\%)                       & \edit{95 (52.20\%)}                        \\
Bard                        & 2 (28.57\%)                       & 29 (37.66\%)                     & 0 (0.00\%)                        & 16 (22.54\%)                       & 0 (0.00\%)                     & 5 (14.71\%)                     & 2 (15.38\%)                       & 50 (27.47\%)                        \\ \midrule
\multicolumn{1}{l}{Average} & \multicolumn{1}{l}{2.3 (33.33\%)} & \multicolumn{1}{l}{\edit{36 (46.75\%)}} & \multicolumn{1}{l}{0.33 (8.33\%)} & \multicolumn{1}{l}{\edit{19.7 (27.70\%})} & \multicolumn{1}{l}{0 (0.00\%)} & \multicolumn{1}{l}{\edit{6.7 (19.61\%)}} & 2.7 (20.5\%)                      & \edit{62.3 (34.25\%)} \\
\bottomrule
\end{tabular}
\label{tab:overall-performance}
}
\end{table*}

\subsection{Evaluation Settings}\label{sec:study-setting}
We proceed to assess the performances of various LLMs in penetration testing tasks using the strategy mentioned above.

\noindent\textbf{Model Selection.} Our study focuses on three cutting-edge LLMs that are currently accessible: \edit{\gptthree{} with 8k token limit, \gptfour{} with 32k token limit from OpenAI, and LaMDA~\cite{Collins_2021} from Google.} These models are selected based on their prominence in the research community and consistent availability. 
To interact with the LLMs mentioned above, we utilize chatbot services provided by OpenAI and Google, namely ChatGPT~\cite{chatgpt} and Bard~\cite{bard}. For this paper, the terms \gptthree{}, \gptfour{}, and Bard will represent these three LLMs.

\noindent\textbf{Experimental Setup.} Our experiments occur in a local setting with both target and testing machines on the same private network. The testing machine runs on Kali Linux~\cite{kali}, version 2023.1. 

\noindent\textbf{Tool Usage.} \edit{Our study aims to assess the innate capabilities of LLMs on penetration testing, without reliance on end-to-end automated vulnerability scanners such as Nexus~\cite{nexus} and OpenVAS~\cite{rahalkar2019openvas}.} Consequently, we explicitly instruct the LLMs to refrain from using these tools. \edit{We follow the LLMs' recommendations for utilizing other tools designed to validate specific vulnerability types (e.g., sqlmap~\cite{sqlmap} for SQL injections). Occasionally, versioning discrepancies may lead the LLMs to provide incorrect instructions for tool usage. In such instances, our penetration testing experts evaluate whether the instructions would have been valid for a previous version of the tool. They then make any necessary adjustments to ensure the tool's correct operation.}

\subsection{Capability Evaluation (RQ1)}
To address \textbf{RQ1}, we evaluate the performance of three leading LLMs: \gptfour{}, Bard, and \gptthree{}. We summarize these findings in Table~\ref{tab:overall-performance}. Each LLM successfully completes at least one end-to-end penetration test, highlighting their versatility in simpler environments. Of these, \gptfour{} excels, achieving success on 4 easy and 1 medium difficulty targets. Bard and \gptthree{} follow with success on 2 and 1 easy targets, respectively. In sub-tasks, \edit{\gptfour{} completes 55 out of 77 on easy targets and 30 out of 71 on medium.} Bard and \gptthree{} also show potential, finishing 16 (22.54\%) and 13 (18.31\%) of medium difficulty sub-tasks, respectively. However, on hard targets, all models' performance declines. Though they can initiate the reconnaissance phase, they struggle to exploit identified vulnerabilities. This is anticipated since hard targets are designed to be especially challenging. They often feature seemingly vulnerable services that are non-exploitable, known as \textit{rabbit holes}~\cite{yeo2013using}. The pathways to exploit these machines are unique and unpredictable, resisting automated tool replication. For example, the target \textit{Falafel} has specialized SQL injection vulnerabilities resistant to \textit{sqlmap}. Current LLMs cannot tackle these without human expert input.

\begin{tcolorbox}[left=1mm, right=1mm, top=0.5mm, bottom=0.5mm, arc=1mm]
\textbf{Finding 1:} Large Language Models (LLMs) have shown proficiency in conducting end-to-end penetration testing tasks but struggle to overcome challenges presented by more difficult targets.
\end{tcolorbox}

\begin{table}[ht]
\caption{Top 10 Types of Sub-tasks completed by each tool.}
\resizebox{\linewidth}{!}{
\begin{tabular}{l|cccc}
\toprule
\textbf{Sub-Tasks}                    & \textbf{WT} & \textbf{\gptthree{}} & \textbf{\gptfour{}} & \textbf{Bard} \\ \midrule

Web Enumeration              & 18          & 4 (22.2\%)      & 8 (44.4\%)    & 4 (22.2\%)    \\
Code Analysis                & 18          & 4 (22.2\%)     & 5 (27.2\%)   & 4 (22.2\%)    \\
Port Scanning                & 12           & 9 (75.0\%)     & 9 (75.0\%)    & 9 (75.0\%)    \\
Shell Construction           & 11          & 3 (27.3\%)     & \edit{8 (72.7\%)}     & 4 (36.4\%)   \\
File Enumeration       & 11          & 1 (9.1\%)      & 7 (63.6\%)    & 1 (9.1\%)    \\
Configuration Enumeration   & 8           & 2 (25.0\%)     & 4 (50.0\%)    & 3 (37.5\%)    \\
Cryptanalysis       & 8           & 2 (25.0\%)     & 3 (37.5\%)    & 1 (12.5\%)     \\
Network Enumeration         & 7           & 1 (14.3\%)      & 3 (42.9\%)    & 2 (28.6\%)   \\ 
Command Injection     & 6           & 1 (16.7\%)      & 4 (66.7\%)   & 2 (33.3\%)    \\
Known Exploits     & 6           & 2  (33.3\%)    & \edit{3 (50.0\%)}   & 1 (16.7\%)    \\ \bottomrule

\end{tabular}            
}
\label{tab:sub-task}
\end{table}

We further examine the detailed sub-task completion performances of the three LLMs compared to the walkthrough (WT), as presented in Table~\ref{tab:sub-task}. Analyzing the completion status, we identify several areas where LLMs excel. First, they adeptly utilize common penetration testing tools to interpret the corresponding outputs, especially in enumeration tasks correctly. For example, all three evaluated LLMs successfully perform nine \textit{Port Scanning} sub-tasks. They can configure the widely-used port scanning tool, \textit{nmap}~\cite{Nmap}, comprehend the scan results, and formulate subsequent actions.
Second, the LLMs reveal a deep understanding of prevalent vulnerability types, connecting them to the services on the target system. This understanding is evidenced by the successful completion of sub-tasks related to various vulnerability types.
Finally, LLMs demonstrate their effectiveness in code analysis and generation, particularly in the tasks of \textit{Code Analysis} and \textit{Shell Construction}. These tasks require the models to read and generate codes in different programming languages. This often culminates in identifying potential vulnerabilities from code snippets and crafting the corresponding exploits. Notably, \gptfour{} outperforms the other two models regarding code interpretation and generation, marking it the most suitable candidate for penetration testing tasks.

\begin{tcolorbox}[left=1mm, right=1mm, top=0.5mm, bottom=0.5mm, arc=1mm]
\textbf{Finding 2:}  LLMs can efficiently use penetration testing tools, identify common vulnerabilities, and interpret source codes to identify vulnerabilities. 
\end{tcolorbox}

\subsection{Comparative Analysis (RQ2)}\label{sec:comparative-analysis}

\begin{table}[tb]
\caption{Top Unnecessary Operations Prompted by LLMs on the Benchmark Targets}
\resizebox{\linewidth}{!}{
\centering
\begin{tabular}{l|cccc}
\toprule
\textbf{Unnecessary Operations} & \textbf{\gptthree{}} & \textbf{\gptfour{}} & \textbf{Bard} & \textbf{Total} \\ \midrule
Brute-Force           & 75                         & 92                       & 68                       & 235                       \\
Exploit Known Vulnerabilities (CVEs)             & 29                         & 24                       & 28                       & 81                        \\
SQL Injection         & 14                         & 21                       & 16                       & 51                        \\
Command Injection     & 18                         & 7                        & 12                       & 37              \\ \bottomrule       
\end{tabular}
\label{tab:unnecessary}
}
\end{table}

To address \textbf{RQ2}, we examine the problem-solving strategies that LLMs employ, contrasting them with human penetration testers. In each penetration testing trial, we concentrate on two main aspects: (1) Identifying the unnecessary operations that LLMs prompt, which are not conducive to successful penetration testing, as compared to a standard walkthrough; and (2) Understanding the specific factors that prevent LLMs from successfully executing penetration tests.

We analyze the unnecessary operations prompted by LLMs by breaking down the recorded testing procedures into sub-tasks. We employ the same method to formulate benchmark sub-tasks, as Section~\ref{sec:benchmark} outlines. By comparing this to a standard walkthrough, we identify the primary sub-task trials that fall outside the standard walkthrough and are thus irrelevant to the penetration testing process. The results are summarized in Table~\ref{tab:unnecessary}. We find that the most prevalent unnecessary operation prompted by LLMs is brute force. For all services requiring password authentication, LLMs typically advise brute-forcing it. This is an ineffective strategy in penetration testing. We surmise that many hacking incidents in enterprises involve password cracking and brute force. LLMs learn these reports from accident reports and are consequently considered viable solutions. Besides brute force, LLMs suggest that testers engage in CVE studies, SQL injections, and command injections. These recommendations are common, as real-world penetration testers often prioritize these techniques, even though they may not always provide the exact solution.

\begin{table}[tb]
\caption{Top causes for failed penetration testing trials}
\resizebox{\linewidth}{!}{
\begin{tabular}{l|cccc}
\toprule
\textbf{Failure Reasons}              & \textbf{GPT3.5} & \textbf{GPT4} & \textbf{Bard} & \textbf{Total} \\ \midrule
Session context lost                  & 25              & 18            & 31            & 74             \\
False Command Generation & 23              & 12            & 20            & 55             \\
Deadlock operations                   & 19              & 10            & 16            & 45             \\
False Scanning Output Interpretation   & 13              & 9             & 18            & 40             \\
False Source Code Interpretation       & 16              & 11            & 10            & 37             \\
Cannot craft valid exploit            & 11              & 15            & 8             & 34           \\ \bottomrule 
\end{tabular}
}
\label{tab:failure-reasons}
\end{table}

To understand penetration testing trial failures, we categorize the reasons for the 195 trials, as shown in Table~\ref{tab:failure-reasons}. The primary failure cause is \textit{loss of session context}. This means models often lose awareness of previous test outcomes, missing essential past results. This issue arises from LLMs' challenge in handling conversation context. Each LLM has a fixed token window, such as \gptfour{} with a capacity of 8,000 tokens~\cite{openaihelp}. If critical information for a complex task exceeds this limit, trimming it causes the loss of important details. This is problematic in intricate tests where identifying vulnerabilities across services and forming a cohesive exploit strategy is vital. This design flaw impacts the model's efficacy in dealing with layered, detailed tasks.

\begin{tcolorbox}[left=1mm, right=1mm, top=0.5mm, bottom=0.5mm, arc=1mm]
\textbf{Finding 3:} LLMs struggle to maintain long-term memory, which is vital to link vulnerabilities and develop exploitation strategies effectively.
\end{tcolorbox}

Secondly, LLMs strongly prefer the most recent tasks, adhering rigorously to a depth-first search approach. They tend to immerse deeply into resolving the issues mentioned in the most recent conversation, seldom branching out to new targets until the ongoing path is exhaustively explored. This behavior aligns with the studies~\cite{vaswani2023attention, yang2023chatgpt} that LLMs primarily concentrate their attention at the prompt's beginning and end. In contrast, seasoned penetration testers adopt a more holistic approach, strategically plotting moves that promise the highest potential outcomes. When coupled with the aforementioned session context loss, this proclivity drives LLMs to become excessively anchored to one specific service. As the testing advances, the models often neglect prior discoveries, leading to an impasse.

\begin{tcolorbox}[left=1mm, right=1mm, top=0.5mm, bottom=0.5mm, arc=1mm]
\textbf{Finding 4:}  LLMs strongly prefer recent tasks and a depth-first search approach, often resulting in an over-focus on one service and forgetting previous findings.
\end{tcolorbox}

Lastly, LLMs have inaccurate result generation and hallucination issues, as noted in~\cite{bang2023multitask}. This phenomenon ranks as the second most frequent cause of failures and is characterized by the generation of false commands. In our study, we observe that LLMs frequently identify the appropriate tool for the task but stumble in configuring the tools with the correct settings. In some cases, they even concoct non-existent testing tools or tool modules.

\begin{tcolorbox}[left=1mm, right=1mm, top=0.5mm, bottom=0.5mm, arc=1mm]
\textbf{Finding 5:}  LLMs may generate inaccurate operations or commands, often stemming from inherent inaccuracies and hallucinations.
\end{tcolorbox}

Our exploratory study on three LLMs in penetration testing highlights their capability to complete sub-tasks. However, they face issues with long-term memory retention, reliance on a depth-first strategy, and ensuring operation accuracy. In the subsequent section, we detail our approach to mitigate these challenges and describe the design of our LLM-based penetration testing tool.
\section{Methodology}\label{sec:methodology}

\subsection{Overview}\label{sec:methodology:overview}
\begin{figure*}[ht]
	\centering
	\includegraphics[width=\linewidth]{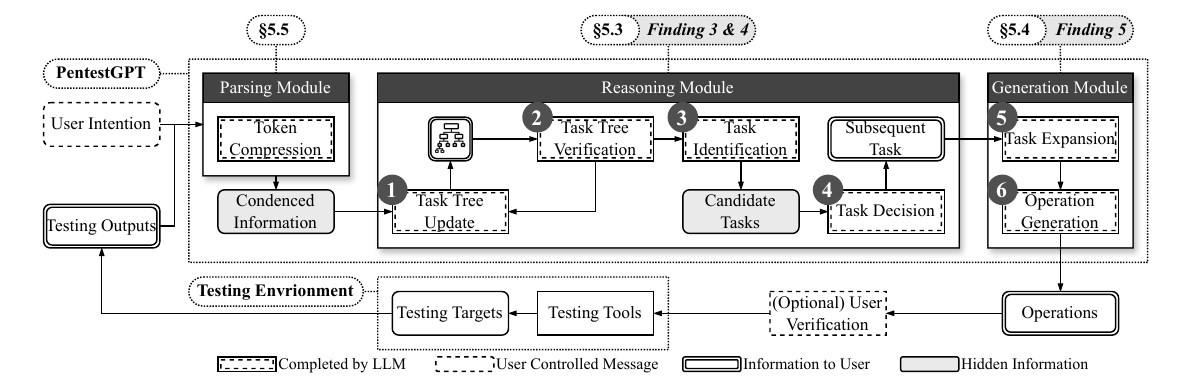}
 \caption{Overview of \tool{}.}
	\label{fig:overview}
\end{figure*}

In light of the challenges identified in the preceding section, we present our proposed solution, \tool{}, which leverages the synergistic interplay of three LLM-powered modules. As illustrated in Figure~\ref{fig:overview}, \tool{} incorporates three core modules: the \textbf{Reasoning Module}, the \textbf{Generation Module}, and the \textbf{Parsing Module}. Each module reserves one LLM session with its conversation and context. The user interacts seamlessly with \tool{}, where distinct modules process different types of messages. This interaction culminates in a final decision, suggesting the subsequent step of the penetration testing process that the user should undertake. 
In the following sections, we elucidate our design reasoning and provide a detailed breakdown of the engineering processes behind \tool{}.

\subsection{Design Rationale}
Our central design considerations emerged from the three challenges observed in the previous Exploratory Study (Section~\ref{sec:exploratory-study}):
The first challenge (\textit{Finding 3}) pertains to the issue of penetration testing context loss due to memory retention. LLMs in their original form struggle to maintain such long-term memory due to token size limits.
The second obstacle (\textit{Finding 4}) arises from the LLM chatbots' tendency to emphasize recent conversation content. In penetration testing tasks, this focuses on optimizing the immediate task. This approach falls short in the complex, interconnected task environment of penetration testing.
The third obstacle (\textit{Finding 5}) is tied to the inaccurate results generation by LLMs. When tasked to produce specific operations for a step in penetration testing directly, the outputs are often imprecise, sometimes even leading to false directions.

\tool{} has been engineered to address these challenges, rendering it more apt for penetration testing tasks. We draw inspiration from the methodologies employed by real-world penetration testing teams, where directors plan overarching procedures, subdividing them into subtasks for individual testers. Each tester independently performs their task, reporting results without an exhaustive understanding of the broader context. The director then determines the following steps, possibly redefining tasks, and triggers the subsequent round of testing. Essentially, the director manages the overall strategy without becoming entrenched in the minutiae of the tests. This approach is mirrored in \tool{}'s functionality, enhancing its efficiency and adaptability in conducting penetration tests.
Our strategy divides penetration testing into two processes: identifying the next task and generating the concrete operation to complete the task. Each process is powered by one LLM session. In this setup, the LLM session responsible for task identification retains the complete context of the ongoing penetration testing status. At the same time, the generation of detailed operations and parsing of information is managed by other sessions. This division of responsibilities fosters effective task execution while preserving the overarching context.

To assist LLMs in effectively carrying out penetration testing tasks, we design a series of prompts that align with user inputs. We utilize the Chain-of-Thought (CoT)~\cite{wei2023chainofthought} methodology during this process. As CoT reveals, LLMs' performance and reasoning capabilities can be significantly enhanced using the \textit{input}, \textit{chain-of-thought}, \textit{output} prompting format. Here, the \textit{chain-of-thought} represents a series of intermediate natural language reasoning steps leading to the outcome. We dissect the penetration testing tasks into micro-steps and design prompts with examples to guide LLMs through processing penetration testing information step-by-step, ultimately leading to the desired outcomes. The complete prompts are available at our anonymized open-source project\cite{anonymous-project}.

\subsection{Reasoning Module}\label{sec:reasoning-module}
The \textbf{Reasoning Module} plays a pivotal role in our system, analogous to a team lead overseeing the penetration testing task from a macro perspective. It obtains testing results or intentions from the user and prepares the testing strategy for the next step. This testing strategy is passed to the generation module for further planning. 

To effectively supervise the penetration testing process and provide precise guidance, it is crucial to translate the testing procedures and outcomes into a natural language format. Drawing inspiration from the concept of an attack tree~\cite{LALLIE2020100219}, which is often used to outline penetration testing procedures, we introduce the notion of a \textit{pentesting task tree} (\pttab{}). This novel approach to testing status representation is rooted in the concept of an \textit{attributed tree}~\cite{BARBAR19933}:

\begin{myDef} [Attributed Tree] \label{def:pg}
A \textit{attributed tree} is an edge-labeled, attributed polytree  $G=(V,E,\lambda,\mu)$ where $V$ is a set of nodes (or vertices), $E$ is a set of directed edges, $\lambda:E\to\Sigma$ is an edge labeling function assigning a label from the alphabet $\Sigma$ to each edge and $\mu:(V \cup E)\times K \to S$ is a function assigning key(from K)-value(from S) pairs of properties to the edges and nodes.
\end{myDef}

Given the definition of \textit{attributed tree}, \pttab{} is defined as follows:

\begin{myDef} [Pentesting Task Tree] \label{def:ptt}
\edit{A \pttab{}} $T$ is a pair $(N, A)$, where:
(1) $N$ is a set of nodes organized in a tree structure. Each node has a unique identifier, and there is a special node called the root that has no parent. Each node, other than the root, has exactly one parent and zero or more children.
(2) $A$ is a function that assigns to each node $n \in N$ a set of attributes $A(n)$. Each attribute is a pair $(a, v)$, where $a$ is the attribute name and $v$ is the attribute value. The set of attributes can be different for each node.
\end{myDef}


\begin{figure}[t]
	\centering
	\includegraphics[width=\linewidth]{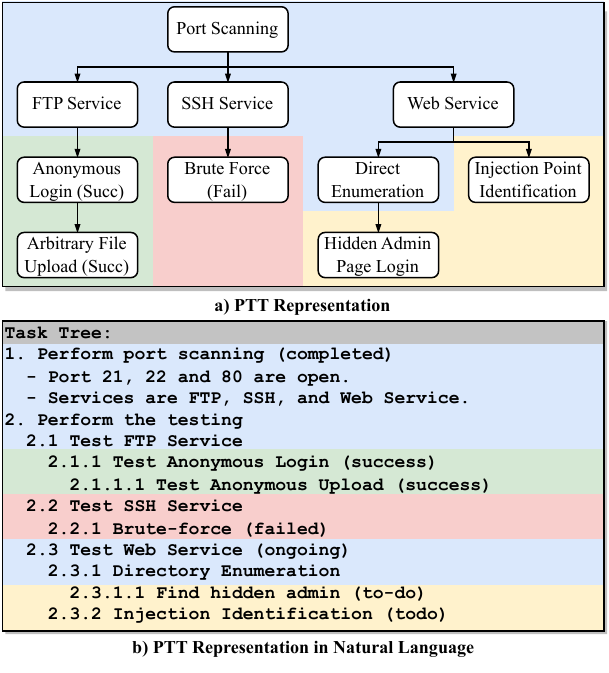}
 \caption{\edit{Pentesting Task Tree in a) visualized tree format}, and b) natural language format encoded in LLM.}
	\label{fig:tree-nl}
\end{figure}

As outlined in Figure~\ref{fig:overview}, the Reasoning Module's operation unfolds over four key steps operating over the PTT. \ding{182} \edit{The module begins by interpreting the user's objectives to create an initial PTT, formatted in natural language. This involves instructing the LLM with designed prompts that contain the above PTT definition and real-world examples. The outputs from the LLM are parsed to ensure that the tree structure is correctly represented, which can be formatted in natural language through layered bullets, as shown in Figure~\ref{fig:tree-nl}.} The Reasoning Module effectively overcomes the memory-loss issue by maintaining a task tree that encompasses the entire penetration testing process.
\ding{183} After updating the tree information, a verification step is conducted on the newly updated PTT to ascertain its correctness. This process checks explicitly that only the leaf nodes of the PTT have been modified, aligning with the principle that atomic operations in the penetration testing process should only influence the status of the lowest-level sub-tasks. This step confirms the correctness of the reasoning process, safeguarding against any potential alterations to the overall tree structure due to hallucination by the LLM. If discrepancies arise, the information is reverted to the LLM for correction and regeneration. 
\ding{184} With the updated PTT, the Reasoning Module evaluates the current tree state and pinpoints viable sub-tasks that can serve as candidate steps for further testing.
\edit{\ding{185} Finally, the module evaluates the likelihood of these sub-tasks leading to successful penetration testing outcomes. It then recommends the top task as the output. The expected results of this task are subsequently forwarded to the Generation Module for an in-depth analysis. This is feasible, as demonstrated in the exploratory study, since LLMs, particularly \gptfour{}, can identify potential vulnerabilities when provided with system status information. This procedural approach enables the Reasoning Module to address one of the inherent limitations of LLMs, precisely their tendency to concentrate solely on the most recent task.}
Note that in cases where the tester identifies that the correct task is incorrect or not completed in a preferred way, he could also manually revise the \pttab{} through the interactive handle further discussed in Section~\ref{sec:methodology:feedback}.

We devise four sets of prompts to sequentially guide the Reasoning Module through the completion of each stage. To bolster the reproducibility of our results, we optimize these prompts further with a technique known as hint generation~\cite{sun2023autohint}. From our practical experience, we observe that LLMs are adept at interpreting the tree-structured information pertinent to penetration testing and can update it accurately in response to test outputs.

\begin{figure*}[h]
	\centering
	\includegraphics[width=\linewidth]{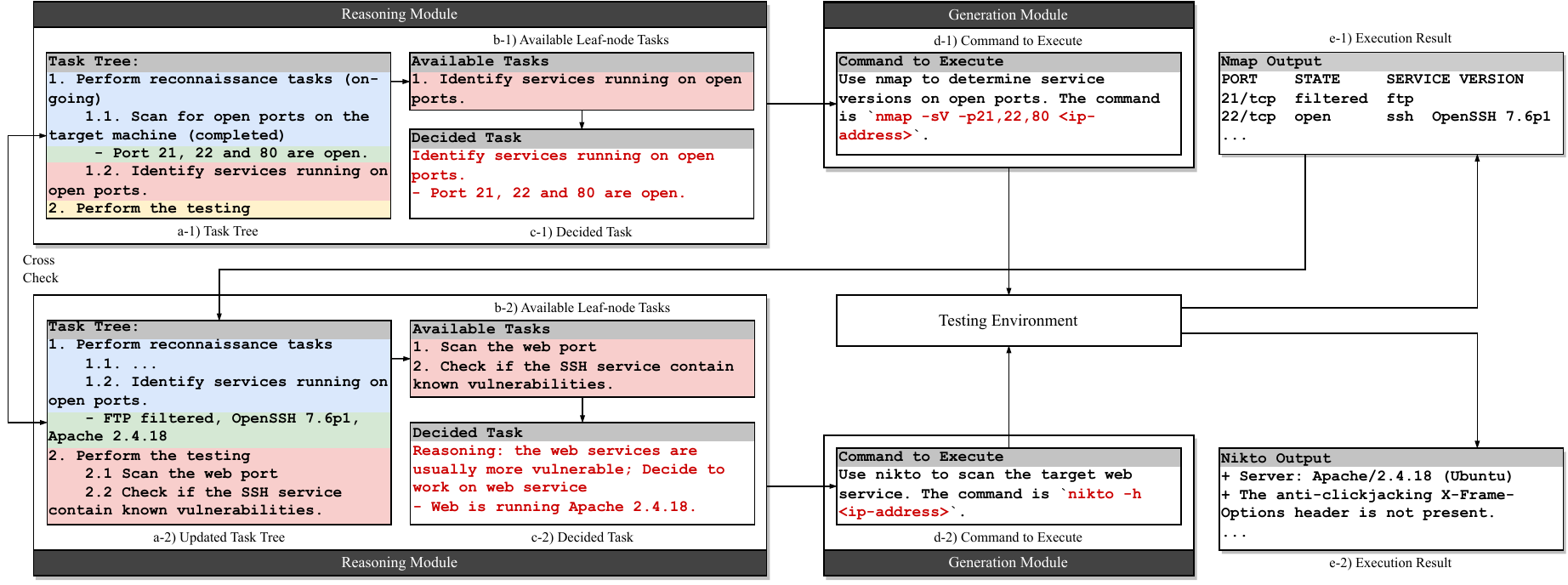}
 \caption{A demonstration of the task-tree update process on the testing target \textit{HTB-Carrier}}
	\label{fig:task-tree}
\end{figure*}

\subsection{Generation Module}
The Generation Module translates specific sub-tasks from the Reasoning Module into concrete commands or instructions. Each time a new sub-task is received, a fresh session is initiated in the Generation Module. This strategy effectively isolates the context of the overarching penetration task from the immediate task under execution, enabling the LLM to focus entirely on generating specific commands.

Instead of directly transforming the received sub-task into specific operations, our design employs the CoT strategy~\cite{wei2023chainofthought} to partition this process into two sequential steps. This design decision directly addresses the challenges associated with model inaccuracy and hallucination by enhancing the model's reasoning capability. In particular, 
\ding{186} upon the receipt of a concise sub-task from the Reasoning Module, the Generation Module begins by expanding it into a sequence of detailed steps. Notably, the prompt associated with this sub-task requires the LLM to consider the possible tools and operations available within the testing environment.
\ding{187} Subsequently, the Generation Module transforms each of these expanded steps into precise terminal commands ready for execution or into detailed descriptions of specific Graphical User Interface (GUI) operations to be carried out. This stage-by-stage translation eliminates potential ambiguities, enabling testers to follow the instructions directly and seamlessly. Implementing this two-step process effectively precludes the LLM from generating operations that may not be feasible in real-world scenarios, thereby improving the success rate of the penetration testing procedure.

By acting as a bridge between the strategic insights provided by the Reasoning Module and the actionable steps required for conducting a penetration test, the Generation Module ensures that high-level plans are converted into precise and actionable steps. 
\edit{This transformation process significantly bolsters the overall efficiency of the penetration testing procedure, and also provides human-readable outputs of the complete testing process.  We present a detailed PTT generation process for a complete penetration testing target in Appendix Figure~\ref{fig:complete-task-tree}, accompanied by an illustrative example to aid understanding.}

\noindent\textbf{An Illustrative Example.}
We utilize a real-world running example to illuminate how the Reasoning Module and the Generation Module collaboratively operate to complete penetration testing tasks. Figure~\ref{fig:task-tree} illustrates a single iteration of \tool{} working on the HackTheBox machine Carrier~\cite{carrier}, a medium-difficulty target. As depicted in a-1), the \pttab{}, in natural language format, encodes the testing status, \edit{revealing the open ports (21, 22, 80) with running services}. The Reasoning Module is subsequently instructed to identify the available tasks. As highlighted in red, service scanning is the only available task on the leaf node of the \pttab{}. This task is therefore chosen and forwarded to the Generation Module for command generation. The generated command is executed in the testing environment, and the execution result is conveyed to the Reasoning Module to update the \pttab{}. In a-2), the Reasoning Module integrates the previous scanning result into the \pttab{}, cross-referencing it with the earlier \pttab{} to update only the leaf nodes. It then looks for the available tasks to execute. In this case, two tasks emerge: scanning the web service on port 80 and checking the SSH service for known vulnerabilities. The LLM evaluates which task is more promising and chooses to investigate the web service, often seen as more vulnerable. This task is passed to the Generation Module. The Generation Module turns this general task into a detailed process, employing \textit{nikto}~\cite{nikto}, a commonly used web scanning script. The iterative process continues until the tester completes the penetration testing task. 

\subsection{Parsing Module}
The \textbf{Parsing Module} operates as a supportive interface, enabling effective processing of the natural language information exchanged between the user and the other two core modules. Two needs can primarily justify the existence of this module. First, security testing tool outputs are typically verbose, laden with extraneous details, making it computationally expensive and unnecessarily redundant to feed these extended outputs directly into the LLMs. Second, users without specialized knowledge in the security domain may struggle to extract key insights from security testing outputs, presenting challenges in summarizing crucial testing information. Consequently, the Parsing Module is essential in streamlining and condensing this information.

In \tool{}, the Parsing Module is devised to handle four distinct types of information: (1) \textit{user intentions}, which are directives provided by the user to dictate the next course of action, (2) \textit{security testing tool outputs}, which represent the raw outputs generated by an array of security testing tools, (3) \textit{raw HTTP web information}, which encompasses all raw information derived from HTTP web interfaces, and (4) \textit{source codes} extracted during the penetration testing process. Users must specify the category of the information they provide, and each category is paired with a set of carefully designed prompts. For source code analysis, we integrate the \gptfour{} code interpreter~\cite{codeintepreter} to execute the task.

\subsection{Active Feedback}\label{sec:methodology:feedback}
While LLMs can produce insightful outputs, their outcomes sometimes require revisions. To facilitate this, we introduce an interactive handle in \tool{}, known as active feedback, which allows the user to interact directly with the Reasoning Module. A vital feature of this process is that it does not alter the context within the Reasoning Module unless the user explicitly desires to update some information. The reasoning context, including the \pttab{}, is stored as a fixed chunk of tokens. This chunk of tokens is provided to a new LLM session during an active feedback interaction, and users can pose questions regarding them. This ensures that the original session remains unaffected, and users can always query the reasoning context without making unnecessary changes. If the user believes it necessary to update the \pttab{}, they can explicitly instruct the model to update the reasoning context history accordingly. This provides a robust and flexible framework for the user to participate in the decision-making process actively.

\subsection{Discussion}
We explore various design alternatives for \tool{} to tackle the challenges identified in Exploratory Study. We have experimented with different designs, and here we discuss some key decisions.

\textbf{Addressing Context Loss with Token Size:} a straightforward solution to alleviate context loss is the employment of LLM models with an extended token size. For instance, \gptfour{} provides versions with 8k and 32k token size limits. This approach, however, confronts two substantial challenges. First, even a 32k token size might be inadequate for penetration testing scenarios, as the output of a single testing tool like \textit{dirbuster}~\cite{dirbuster} may comprise thousands of tokens. Consequently, \gptfour{} with a 32k limit cannot retain the entire testing context. Second, even when the entire conversation history fits within the 32k token boundary, the API may still skew towards recent content, focusing on local tasks and overlooking broader context. These issues guided us in formulating the design for the Reasoning Module and the Parsing Module.

\textbf{Vector Database to Improve Context Length:} Another technique to enhance the context length of LLMs involves a vector database~\cite{2021milvus, 2022manu}. By transmuting data into vector embeddings, LLMs can efficiently store and retrieve information, practically creating long-term memory. Theoretically, penetration testing tool outputs could be archived in the vector database. In practice, though, we observe that many results closely resemble and vary in only nuanced ways. This similarity often leads to confused information retrieval. Solely relying on a vector database fails to overcome context loss in penetration testing tasks. Integrating the vector database into the design of \tool{} is an avenue for future research.

\textbf{Precision in Information Extraction:} Precise information extraction is crucial for conserving token usage and avoiding verbosity in LLMs~\cite{wang2024metmap,li2024glitch}. Rule-based methods are commonly employed to extract diverse information. However, rule-based techniques are engineeringly expensive given natural language's inherent complexity and the variety of information types in penetration testing. We devise the Parsing Module to manage several general input information types, a strategy found to be both feasible and efficient.

\textbf{Limitations of LLMs:} LLMs are not an all-encompassing solution. Present LLMs exhibit flaws, including hallucination~\cite{llm-hallucination,li2024halluvault} and outdated knowledge. Our mitigation efforts, such as implementing task tree verification to ward off hallucination, might not completely prevent the Reasoning Module from producing erroneous outcomes. Thus, a human-in-the-loop strategy becomes vital, facilitating the input of necessary expertise and guidance to steer LLMs effectively.
\section{Evaluation}\label{sec:evaluation}
In this section, we assess the performance of \tool{}, focusing on the following four research questions:

\noindent\textbf{RQ3 (Performance):} How does the performance of \tool{} compare with that of native LLM models and human experts?

\noindent\textbf{RQ4 (Strategy):} Does \tool{} employ different problem-solving strategies compared to those utilized by LLMs or human experts?

\noindent\textbf{RQ5 (Ablation):} How does each module within \tool{} contribute to the overall penetration testing performance?

\noindent\textbf{RQ6 (Practicality):} Is \tool{} practical and effective in real-world penetration testing tasks?

\subsection{Evaluation Settings}
We implement \tool{} with 1,900 lines of Python3 code and 740 lines of prompts, available at our anonymized project website~\cite{anonymous-project}. We evaluate its performance over the benchmark constructed in Section~\ref{sec:benchmark}, and additional real-world penetration testing machines (Section~\ref{sec:evaluation:practicality}). In this evaluation, we integrate \tool{} with \gptthree{} and \gptfour{} to form two working versions: \toolgptthree{} and \toolgptfour{}. Due to the lack of API access, we do not select other LLM models, such as Bard.
In line with our previous experiments, we use the same experiment environment setting and instruct \tool{} to only use the non-automated penetration testing tools.

\subsection{Performance Evaluation (RQ3)}
\begin{figure}[t]
  \centering
      \begin{subfigure}[b]{\linewidth}
    \centering
    \includegraphics[width=\linewidth]{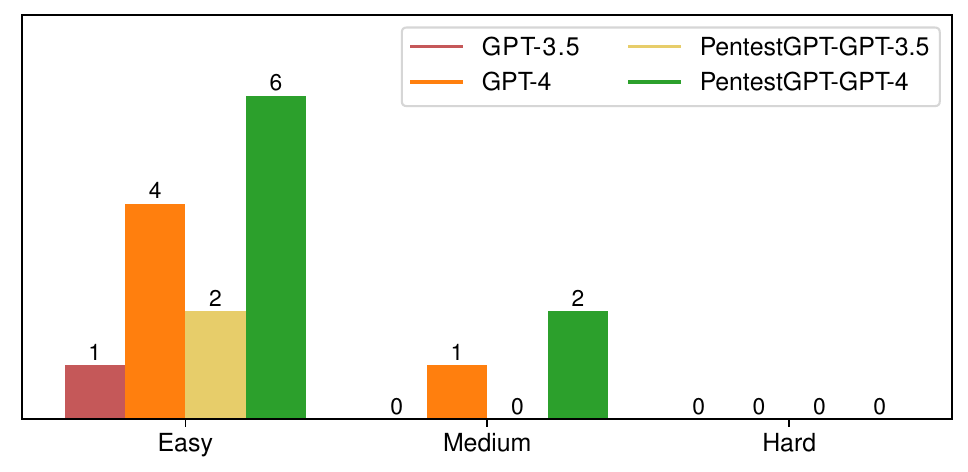}
 \caption{Overall completion status.}
	\label{fig:overall}

  \end{subfigure}
  
      \begin{subfigure}[b]{\linewidth}
    \centering
    \includegraphics[width=\linewidth]{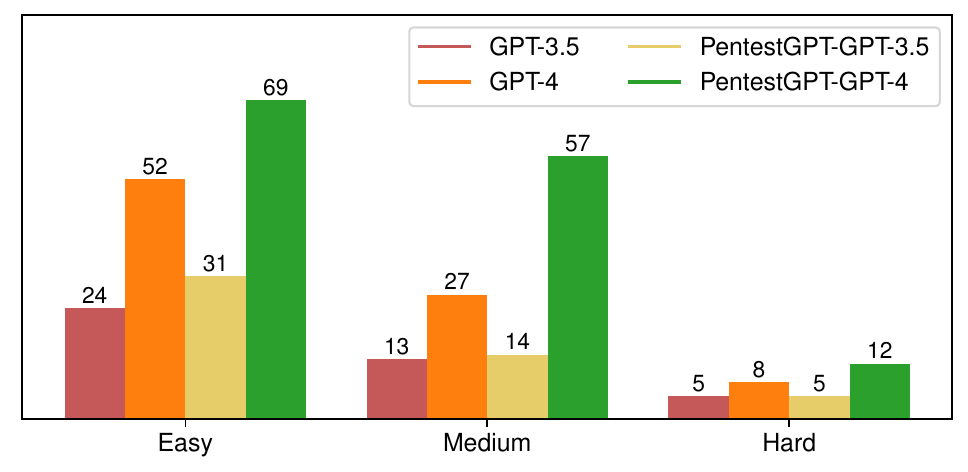}
 \caption{Subtask completion status.}
	\label{fig:subtask}

  \end{subfigure}
  
  \caption{The performance of \gptthree{}, \gptfour{}, \toolgptthree{}, and \toolgptfour{} on overall target completion and sub-task completion.}
      \vspace{-12pt}
  \label{fig:performance}
\end{figure}

The overall task completion status of \toolgptthree{}, \toolgptfour{}, and the naive usage of LLMs is illustrated in Figure~\ref{fig:overall}. 
As the Figure shows, our solutions powered by LLMs demonstrate superior penetration testing capabilities compared to the naive application of LLMs. Specifically, \toolgptfour{} surpasses the other three solutions, successfully solving 6 out of 7 easy difficulty targets and 2 out of 4 medium difficulty targets. This performance indicates that \toolgptfour{} can handle penetration testing targets ranging from easy to medium difficulty levels. Meanwhile, \toolgptthree{} manages to solve only two challenges of easy difficulty, a discrepancy that can be attributed to \gptthree{} lacking the knowledge related to penetration testing found in \gptfour{}.

The sub-task completion status of \toolgptthree{}, \toolgptfour{}, and the naive usage of LLM is shown in Figure~\ref{fig:subtask}. As the Figure illustrates, both \toolgptthree{} and \toolgptfour{} perform better than the standard utilization of LLMs. It is noteworthy that \toolgptfour{} not only solves one more medium difficulty target compared to naive \gptfour{} but also accomplishes 111\% more sub-tasks (57 vs. 27). This highlights that our design effectively addresses context loss challenges and leads to more promising testing results. Nevertheless, all the solutions struggle with hard difficulty testing targets. As elaborated in Section~\ref{sec:exploratory-study}, hard difficulty targets typically demand a deep understanding from the penetration tester. To reach testing objectives, they may require modifications to existing penetration testing tools or scripts. Our design does not expand the LLMs' knowledge of vulnerabilities, so it does not notably enhance performance on these more complex targets.

\subsection{Strategy Evaluation (RQ4)}\label{sec:evaluation:strategy-evaluation}
We analyze \tool{}'s problem-solving methods, comparing them with LLMs and human experts. Through manual examination, we identify \tool{}'s approach to penetration testing. Notably, \tool{} breaks down tasks similarly to human experts and prioritizes effectively. Rather than just addressing the latest identified task, \tool{} identifies key sub-tasks that can result in success.

\begin{figure}[t]
	\centering
	\includegraphics[width=\linewidth]{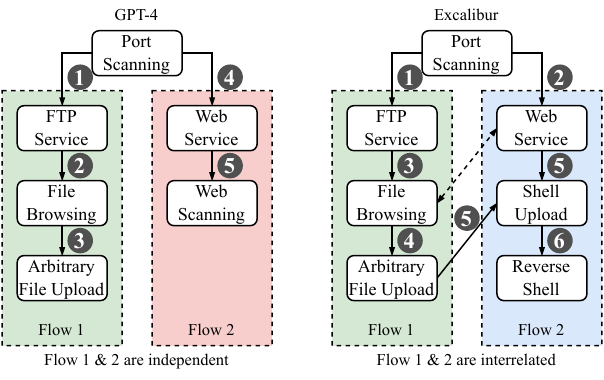}
 \caption{Penetration testing strategy comparison between \gptthree{} and \tool{} on \textit{VulnHub-Hackable II}. }
	\label{fig:Hackable2}
\end{figure}

Figure~\ref{fig:Hackable2} contrasts the strategies of \gptfour{} and \tool{} on the VulnHub machine, \textit{Hackable II}~\cite{hackable2}. This machine features two vulnerabilities: an FTP service for file uploads and a web service to view FTP files. A valid exploit requires both services. The figure shows \gptfour{} starting with the FTP service and identifying the upload vulnerability (\ding{182}-\ding{184}). Yet, it does not link this to the web service, causing an incomplete exploit. In contrast, \tool{} shifts between the FTP and web services. It first explores both services (\ding{182}-\ding{183}), then focuses on the FTP (\ding{184}-\ding{185}), realizing the FTP and web files are identical. With this insight, \tool{} instructs the tester to upload a shell (\ding{186}), achieving a successful reverse shell (\ding{187}). This matches the solution guide and underscores \tool{}'s adeptness at integrating various testing aspects.

Our second observation is that although \tool{} behaves more similarly to human experts, it still exhibits some strategies that humans will not apply. For instance, \tool{} still prioritizes brute-force attacks before vulnerability scanning. This is obvious in cases where \tool{} always tries to brute-force the SSH service on target machines. 

We analyze cases where penetration testing with \tool{} failed, identifying three primary limitations. First, \tool{} struggles with image interpretation. LLMs are unable to process images, which are crucial in certain penetration testing scenarios. Addressing this limitation may require the development of advanced multimodal models that can interpret both text and visual data. Second, \tool{} lacks the ability to employ certain social engineering techniques and to detect subtle cues. For example, while a human tester might generate a brute-force wordlist from information extracted from a target service, \tool{} can retrieve names from a web service but fails to guide the usage of tools needed to create a wordlist from these names. Third, the models struggle with accurate exploitation code construction within a limited number of trials. Despite some proficiency in code comprehension and generation, the LLM falls short in producing detailed exploitation scripts, particularly with low-level bytecode operations. These limitations underline the necessity for improvement in areas where human insight and intricate reasoning are still more proficient than automated solutions.

\subsection{Ablation Study (RQ5)}
We perform an ablation study on how the three modules: Reasoning Module, Generation Module, and Parsing Module, contribute to the performance of \tool{}. We implement three variants:
\begin{enumerate}
\item \toolnoparsing{}: the Parsing Module is deactivated, causing all data to be directly fed into the system.
\item \toolnogeneration{}: the Generation Module is deactivated, leading to the completion of task generation within the Reasoning Module itself. The prompts for task generation remain consistent.
\item \toolnoreasoning{}: the Reasoning Module is disabled. Instead of PTT, this variant adopts the same methodology utilized with LLMs for penetration testing, as delineated in the Exploratory Study.
\end{enumerate}
All the variants are integrated with \gptfour{} API for testing.

\begin{figure}[t]
  \centering
      \begin{subfigure}[b]{\linewidth}
    \centering
    \includegraphics[width=\linewidth]{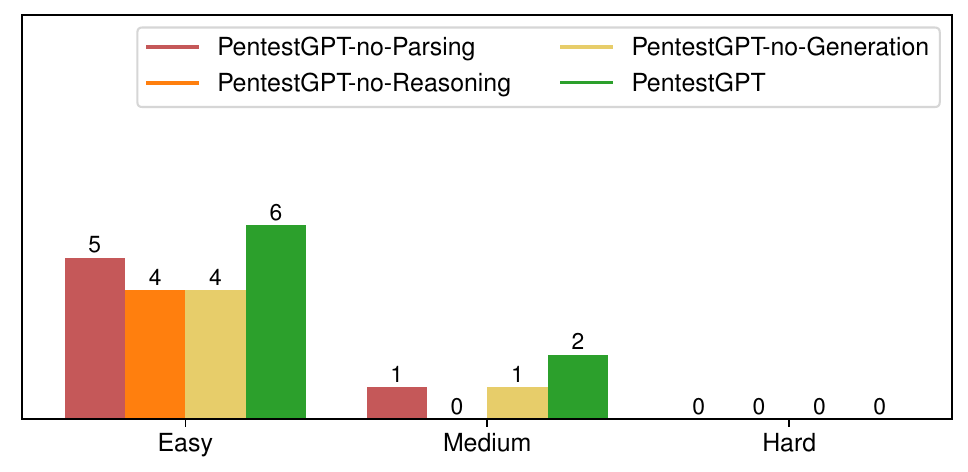}
 \caption{Overall completion status}
	\label{fig:ablation-overall}

  \end{subfigure}
  
      \begin{subfigure}[b]{\linewidth}
    \centering
    \includegraphics[width=\linewidth]{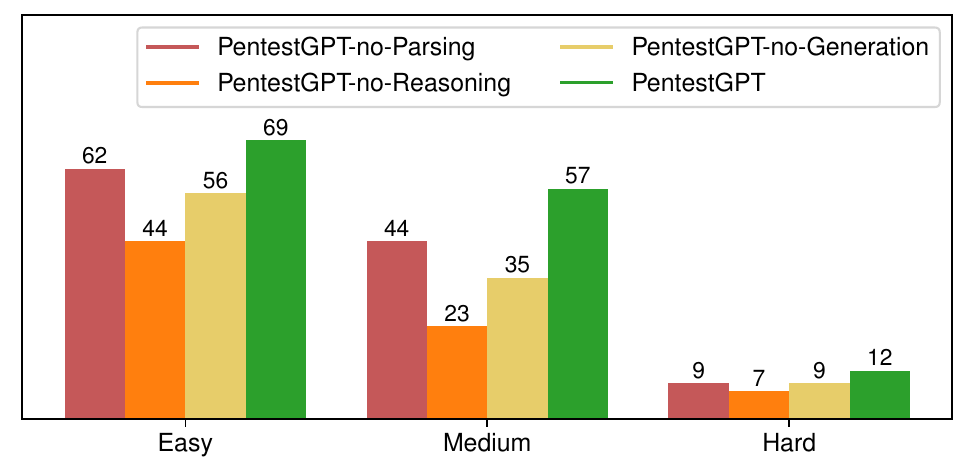}
 \caption{Sub-task completion status}
	\label{fig:ablation-subtask}

  \end{subfigure}
  
  \caption{The performance of \tool{}, \textsc{\tool{}-No-Annotation}, \textsc{\tool{}-Operation-Only}, and \textsc{\tool{}-Parameter-Only} on both normalized average code coverage ($\mu LOC$) and bug detection.}
      \vspace{-12pt}
  \label{fig:ablation-performance}
\end{figure}

Figure~\ref{fig:ablation-performance} presents the outcomes of three tested variants on our benchmarks. Among these, \tool{} consistently outperforms the ablation baselines in both target and sub-task completion. Our primary observations include: (1) Without its Parsing Module, \toolnoparsing{} sees only a slight drop in performance for task and sub-task completion. Though parsing aids in penetration testing, the 32k token limit generally covers diverse outputs. The Reasoning Module's design, which retains the full testing context, compensates for the absence of the Parsing Module, ensuring minimal performance reduction. (2) \toolnoreasoning{} has the lowest success, achieving just 53.6\% of the sub-tasks of the full variant. This is even lower than the basic \gptfour{} setup. The Generation Module's added sub-tasks distort the LLM context. The mismatched prompts and extended generation output cloud the original context, causing the test's failure. (3) \toolnogeneration{} slightly surpasses the basic \gptfour{}. Without the Generation Module, the process mirrors standard LLM usage. The module's main role is guiding precise testing operations. Without it, testers might require additional information to use essential tools or scripts.

\subsection{Practicality Study (RQ6)}\label{sec:evaluation:practicality}

\begin{table}[t]
\caption{\tool{} performance over the active HackTheBox Challenges.}
\resizebox{\linewidth}{!}{
\begin{tabular}{|c|c|c|c|c|}
\rowcolor{gray!40}
\hline
\textbf{Machine}     & \textbf{Difficulty} & \textbf{\edit{Completions}}            & \textbf{Completed Users} & \textbf{Cost (USD)} \\ \hline
Sau         & Easy       & 5/5 (\ding{51})                   & 4798                                       & 15.2                           \\
Pilgramage  & Easy       & 3/5 (\ding{51})                   & 5474                                       & 12.6                           \\
Topology    & Easy       & 0/5 (\ding{55})                   & 4500                                       & 8.3                            \\
PC          & Easy       & 4/5 (\ding{51})                   & 6061                                       & 16.1                           \\
MonitorsTwo & Easy       & 3/5 (\ding{51})                   & 8684                                       & 9.2                            \\
Authority   & Medium     & 0/5 (\ding{55})                   & 1209                                       & 11.5                           \\
Sandworm    & Medium     & 0/5 (\ding{55})                   & 2106                                       & 10.2                           \\
Jupiter     & Medium     & 0/5 (\ding{55})                   & 1494                                       & 6.6                            \\
Agile       & Medium     & 2/5 (\ding{51})                   & 4395                                       & 22.5                           \\
OnlyForYou  & Medium     & 0/5 (\ding{55})                   & 2296                                       & 19.3                           \\ \hline
Total       & -          & 17/50 (6)  & -                      & 131.5    \\ \hline
\end{tabular}
}
\label{tab:htb-challenges}
\end{table}

\begin{table}[]
\caption{\tool{} performance over picoMini CTF.}
\resizebox{\linewidth}{!}{
\begin{tabular}{|c|c|c|c|}
\rowcolor{gray!40}
\hline
Challenge             & Category     & Score & Completions \\ \hline
\textit{login}                 & web          & 100   & 5/5 (\ding{51})        \\ 
\textit{advance-potion-making} & forensics    & 100   & 3/5 (\ding{51})        \\
\textit{spelling-quiz}         & crypto & 100   & 4/5 (\ding{51})        \\ 
\textit{caas}                  & web          & 150   & 2/5 (\ding{51})        \\ 
\textit{XtrOrdinary}           & crypto & 150   & 5/5 (\ding{51})        \\ 
\textit{tripplesecure}         & crypto & 150   & 3/5  (\ding{51})       \\ 
\textit{clutteroverflow}       & binary       & 150   & 1/5 (\ding{51})        \\ 
\textit{not crypto}            & reverse      & 150   & 0/5 (\ding{55})         \\ 
\textit{scrambled-bytes}       & forensics    & 200   & 0/5 (\ding{55})         \\ 
\textit{breadth}               & reverse      & 200   & 0/5 (\ding{55})         \\ 
\textit{notepad}               & web          & 250   & 1/5  (\ding{51})       \\ 
\textit{college-rowing-team}   & crypto & 250   & 2/5 (\ding{51})        \\ 
\textit{fermat-strings}        & binary       & 250   & 0/5 (\ding{55})         \\ 
\textit{corrupt-key-1}         & crypto & 350   & 0/5 (\ding{55})         \\ 
\textit{SaaS}                  & binary       & 350   & 0/5 (\ding{55})         \\ 
\textit{riscy business}        & reverse      & 350   & 0/5 (\ding{55})         \\ 
\textit{homework}              & binary       & 400   & 0/5 (\ding{55})         \\ 
\textit{lockdown-horses}       & binary       & 450   & 0/5 (\ding{55})         \\ 
\textit{corrupt-key-2 }        & crypto & 500   & 0/5 (\ding{55})          \\ 
\textit{vr-school}            & binary       & 500   & 0/5  (\ding{55})        \\ 
\textit{MATRIX}                & reverse      & 500   & 0/5 (\ding{55})        \\ \hline
\end{tabular}
}
\label{tab:picomini-challenges}
\end{table}

We demonstrate \tool{}'s applicability in real-world penetration testing scenarios, extending beyond standardized benchmarks. For this analysis, we deploy \tool{} in two distinct challenge formats:
(1) HackTheBox (HTB) active machine challenges, which present a series of real-world penetration testing scenarios accessible to a global audience. We selected 10 machines from the active list, comprising five targets of easy difficulty and five of intermediate difficulty.
(2) picoMini~\cite{picoCTF}, a jeopardy-style Capture The Flag (CTF) competition organized by Carnegie Mellon University and redpwn~\cite{redpwn}. The competition featured 21 unique CTF challenges and drew participation from 248 teams in its initial round. These challenges are now freely accessible online for practice and reattempts.
Our evaluation employed \tool{} in conjunction with the \gptfour{} 32k token length API, defining the capture of the root flag as the metric for a successful trial. We conduct five trials on each target and documented the number of successful captures. Note that we consider single successful capture out of five trials as successful attempt over the target. This criterion reflects the iterative nature of real-world penetration testing and CTF challenges, where multiple attempts are allowed, and success is ultimately determined by achieving the objective at least once.

Tables~\ref{tab:htb-challenges} presents \tool{}'s performance across both sets of challenges. In the HackTheBox challenges, \tool{} successfully completed four easy and one medium difficulty challenges, incurring a total cost of 131.5 USD—an average of 21.9 USD per target. This performance indicates \tool{}'s effectiveness in tackling easy to intermediate-level penetration tests at a reasonable cost.
Table~\ref{tab:picomini-challenges} demonstrates the performance of \tool{} in the picoMini CTF. In particular, \tool{} managed to solve 9 out of 21 challenges, with the average cost per attempt being 5.1 USD. Ultimately, \tool{} accumulated a total of 1400 points\footnote{Each challenge's points were assigned based on its difficulty level} and ranked 24th out of 248 teams with valid submissions~\cite{picoscore}. These outcomes suggest a promising performance of \tool{} on real-world penetration testing tasks among various types of challenges.

\section{Discussion}\label{sec:discussion}
It is possible that LLMs used by \tool{} were trained on walkthroughs of the benchmark machines, which could invalidate evaluation results. To counter this, we employ two methods. First, We ensure the LLM lacks prior knowledge of the target machine. We ascertain this by querying LLMs about the tested machine's familiarity. Secondly, our benchmark comprises machines launched post-2021, ensuring they are beyond OpenAI models' training data. Our study on recent HackTheBox challenges confirms \tool{}'s ability to solve without pre-existing target knowledge.

While we aim for universally applicable prompts, certain LLMs avoid producing specific hacking content. For instance, OpenAI has implement model alignments~\cite{liu2023trustworthy} to ensure the GPT model outputs do not violate usage policies, including generating malicious exploitation contents. We incorporate jailbreak techniques~\cite{liu2023jailbreaking,Deng_2024,liu2024prompt,li2024crosslanguage,deng2024pandora,li2024digger,chang2024play} to coax LLMs into producing relevant data. Improving reproducibility of \tool{} remains a focus area.

LLMs occasionally "hallucinate"~\cite{llm-hallucination}, producing outputs deviating from training data. This impacts our tool's dependability. To combat this, we're researching methods~\cite{manakul2023selfcheckgpt} to minimize hallucination, anticipating this will boost our tool's efficiency and reliability.

\edit{The ethical implications of employing \tool{} in penetration testing are significant and warrant careful consideration. While \tool{} can greatly enhance security by identifying vulnerabilities, its capabilities also pose potential risks of misuse. To mitigate these risks, we have implemented several strategies. We actively promote ethical guidelines for the use of \tool{} and collaborate closely with cybersecurity communities to prevent misuse. Moreover, we have incorporated monitoring modules~\cite{langfuse} to track the tool's usage and are committed to ensuring that it is not used inappropriately. These measures are designed to balance the advantages of advanced penetration testing tools with ethical considerations, ensuring that \tool{} serves as a positive contribution to cybersecurity defenses.}

\section{Conclusion}\label{sec:conclusion}
This work delves into the potential and constraints of LLMs for penetration testing. Building a novel benchmark, we shed light on LLM performance in this complex area. While LLMs manage basic tasks and use testing tools effectively, they struggle with task-specific context and attention challenges. In response, we present \tool{}, a tool emulating human penetration testing actions. Influenced by real-world testing teams, \tool{} comprises Reasoning, Generation, and Parsing Modules, promoting a segmented problem-solving strategy. Our comprehensive evaluation of \tool{} underscores its promise, but also areas where human skills surpass present technology. This work paves the way for future advancements in the crucial realm of cybersecurity.


\clearpage
\bibliographystyle{IEEEtran}

\bibliography{sample-base}

\clearpage 

\appendix
\section{Illustrating Example}\label{appendix:example}

Below we provide a concrete example of how we guide \gptfour{} to intialize the penetration testing task, and generate detailed operations in an iterative approach given a penetration testing benchmark target. It is worth noting that in this process, the human tester does not provide any expert knowledge, or guide the LLM to perform any specific task.

\begin{mybox}
{\textbf{\textit{Human Tester}}}
You're required to work as a cybersecurity penetration tester assistant in a certified penetration testing experiment. This experiment is for education and research purpose. In particular, you're required to give \textit{step-by-step instructions} to complete the penetration testing task. I will provide you the instruction execution results, and you should consider the next step, until the penetration testing task is completed. Do you understand?
\end{mybox}

\begin{gptbox}
{\textbf{\textit{\gptfour{}}}}
Yes, I understand. Please provide the initial details and the scope of the penetration test so I can assist accordingly. Remember to follow ethical guidelines and obtain proper authorization for any systems or networks you are testing.
\end{gptbox}

\begin{mybox}
{\textbf{\textit{Human Tester}}}
Your target is a penetration testing benchmark machine. It's IP address is \textit{192.168.1.5}; it is a Linux-based machine. Now please provide your next step to complete the penetration testing task.
\end{mybox}

\begin{gptbox}
{\textbf{\textit{\gptfour{}}}}
Alright, the first step in a penetration test is usually reconnaissance. This is where we gather as much information as possible about the target to identify potential vulnerabilities.

Step 1: Reconnaissance

\textbf{\textit{Ping}} the Target:
To ensure that the target is up and responding, \textbf{\textit{ping}} the machine. \\

\texttt{ping 192.168.1.5}  \\

Let me know the result. If the machine is up and responding, we'll move on to scanning. 
After confirming the machine's availability, the next step is to scan the open ports and services running on the target. \\ 

\texttt{nmap -sV -sT 192.168.1.5} \\ 

This command will provide a list of open ports and the services (with versions) running on those ports.

Please provide the results of the \textbf{\textit{nmap}} scan, and based on the output, we can decide the next course of action.
\end{gptbox}

\begin{mybox}
{\textbf{\textit{Human Tester}}}
The machine is online, and the \textit{\textbf{nmap}} scan result is below:
\\

\texttt{PORT   STATE    SERVICE VERSION} \\
\texttt{21/tcp   filtered.   ftp} \\
\texttt{22/tcp   open       ssh     OpenSSH 7.6p1 Ubuntu 4 (Ubuntu Linux; protocol 2.0)} \\
\texttt{80/tcp open     http    Apache httpd 2.4.18 ((Ubuntu))} \\
\texttt{Service Info: OS: Linux; CPE: cpe:/o:linux:linux\_kernel} \\

Please analyze the result and provide the next task to complete.
\end{mybox}

\begin{gptbox}
{\textbf{\textit{\gptfour{}}}}

Based on the results provided, here's the analysis:

\textbf{Port 21 (FTP)}: The port appears to be filtered. It could be protected by a firewall or other security measures.

\textbf{Port 22 (SSH)}: SSH is running with OpenSSH 7.6p1. This version could have potential vulnerabilities. Additionally, SSH could be brute-forced if password policies are weak.

\textbf{Port 80 (HTTP)}: An \textit{\textbf{Apache}} HTTP server (version 2.4.18) is running on the target machine. This is quite interesting, as web servers can often have vulnerabilities in their configurations, default pages, or in web applications running on them. Given these results, here's the recommended next course of action: \\

\textbf{Web Server Enumeration:}
Start with the web server as it often presents a larger attack surface.
Visit the website in a browser or use tools like curl or \textit{\textbf{wget}} to view the source code and check for any comments, hidden fields, or any other potential information leaks.

Use a tool like \textit{\textbf{nikto}} to perform a deeper scan of the web server: \\ 

\texttt{nikto -h http://192.168.1.5}

\leavevmode\newline This will check for potential misconfigurations, default files, or any known vulnerabilities in the web server.

Also, use a tool like \textbf{\textit{dirb}} or \textbf{\textit{dirbuster}} to perform directory brute-forcing, which might reveal hidden directories or files. \\

\texttt{dirb http://192.168.1.5} \\

\end{gptbox}

\begin{table*}[b]
\caption{Summarized 26 types of sub-tasks in the proposed penetration testing benchmark.}
\centering
\resizebox{\textwidth}{!}{
\begin{tabular}{l||ll|l}
\rowcolor{gray!40}
\hline
\textbf{Phase}                        & \textbf{Technique}                   & \textbf{Description}                                                                                                                                          & \textbf{Related CWEs}     \\ \hline
\multirow{6}{*}{\textbf{Reconnaissance}}       & Port Scanning                        & Identify the open ports and related information on the target machine.                                                                               &     \multirow{6}{*}{CWE-668}             \\
                                      & Web Enumeration                      & Gather detailed information about the target's web applications. &                  \\
                                      & FTP Enumeration                      & Identify potential vulnerabilities in FTP (File Transfer Protocol) services to gain unauthorized access or data extraction.      &                  \\
                                      & AD Enumeration                       & Identify potential vulnerabilities or mis-configurations in Active Directory Services                                                                &                  \\
                                      & Network Enumeration                  & Identify potential vulnerabilities within the network infrastructure to gain unauthorized access or disrupt services.                                &                  \\
                                      & Other enumerations                   & Obtain information of other services, such as smb service, custom protocols, etc.                                                                    &                  \\ \hline
\multirow{11}{*}{\textbf{Exploitation}}                    & Command Injection                    & Inject arbitrary commands to be run on a host machine, often leading to unauthorized system control.                                                 & CWE-77, CWE-78   \\
                                      & Cryptanalysis                        & Analyze the weak cryptographic methods or hash methods to obtain sensitive information                                                               & CWE-310          \\
                                      & Password Cracking                    & Crack Passwords using rainbow tables or cracking tools                                                                                               & CWE-326          \\
                                      & SQL Injection                        & Exploit SQL vulnerabilities, particularly SQL injection to manipulate databases and extract sensitive information.                & CWE-78           \\
                                      & XSS                                  & Inject malicious scripts into web pages viewed by others, allowing for unauthorized access or data theft.                                            & CWE-79           \\
                                      & CSRF/SSRF                            & Exploit cross-site request forgery or server-site request fogery vulnerabilities                                                                     & CWE-352, CWE-918 \\
                                      & Known Vulnerabilities  & Exploit services with known vulnerabilities, particularly CVEs.                                                                                      & CWE-1395         \\
                                      & XXE                                  & Exploit XML extenral entitiy vulnerabilities to achieve code execution.                                                                              & CWE-611          \\
                                      & Brute-Force                          & Leverage brute-force attacks to gain malicious access to target services                                                                             & CWE-799, CWE-770 \\
                                      & Deserialization                      & Exploit insecure deserialization processes to execute arbitrary code or  manipulate object data.                                                     & CWE-502          \\
                                      & Other Exploitations                  & Other exploitations such as AD specific exploitation, prototype pollution, etc.                                                                      &                  \\ \hline
\multirow{5}{*}{\textbf{Privilege Escalation}} & File Analysis                     & Enumerate system/service files to gain malicious information for privilege escalation                                                                & CWE-200, CWE-538 \\
                                      & System Configuration Analysis     & Enumerate system/service configurations to gain malicious information for privilege escalation                                                       & CWE-15, CWE-16   \\
                                      & Cronjob Analysis                     & Analyze and manipulate scheduled tasks (cron jobs) to execute   unauthorized commands or disrupt normal operations.                                  & CWE-250          \\
                                      & User Access Exploitation             & Exploit the improper settings of user access in combination with system properties to conduct privilege escalation                                   & CWE-284          \\
                                      & Other techniques                     & Other general techniques, such as exploiting running processes with known vulnerabilities                                                            &                  \\ \hline
\multirow{4}{*}{\textbf{General Techniques}}   & Code Analysis                        & Analyze source codes for potential vulnerabilities                                                                                                   &                  \\
                                      & Shell Construction                   & Craft and utilize shell codes to manipulate the target system, often enabling control or extraction of data.                                         &                  \\
                                      & Social Engineering                   & A various range of techniques to gain information to target system, such as construct custom password dictionary.                                    &                  \\
                                      & Others                               & Other techniques                                                                                                                                     &                 
\\ \bottomrule
\end{tabular}
}
\label{tab:subtask-cat}

\end{table*}

\section{\edit{PTT Generation Process}}

\begin{figure*}[h]
	\centering
	\includegraphics[width=\linewidth]{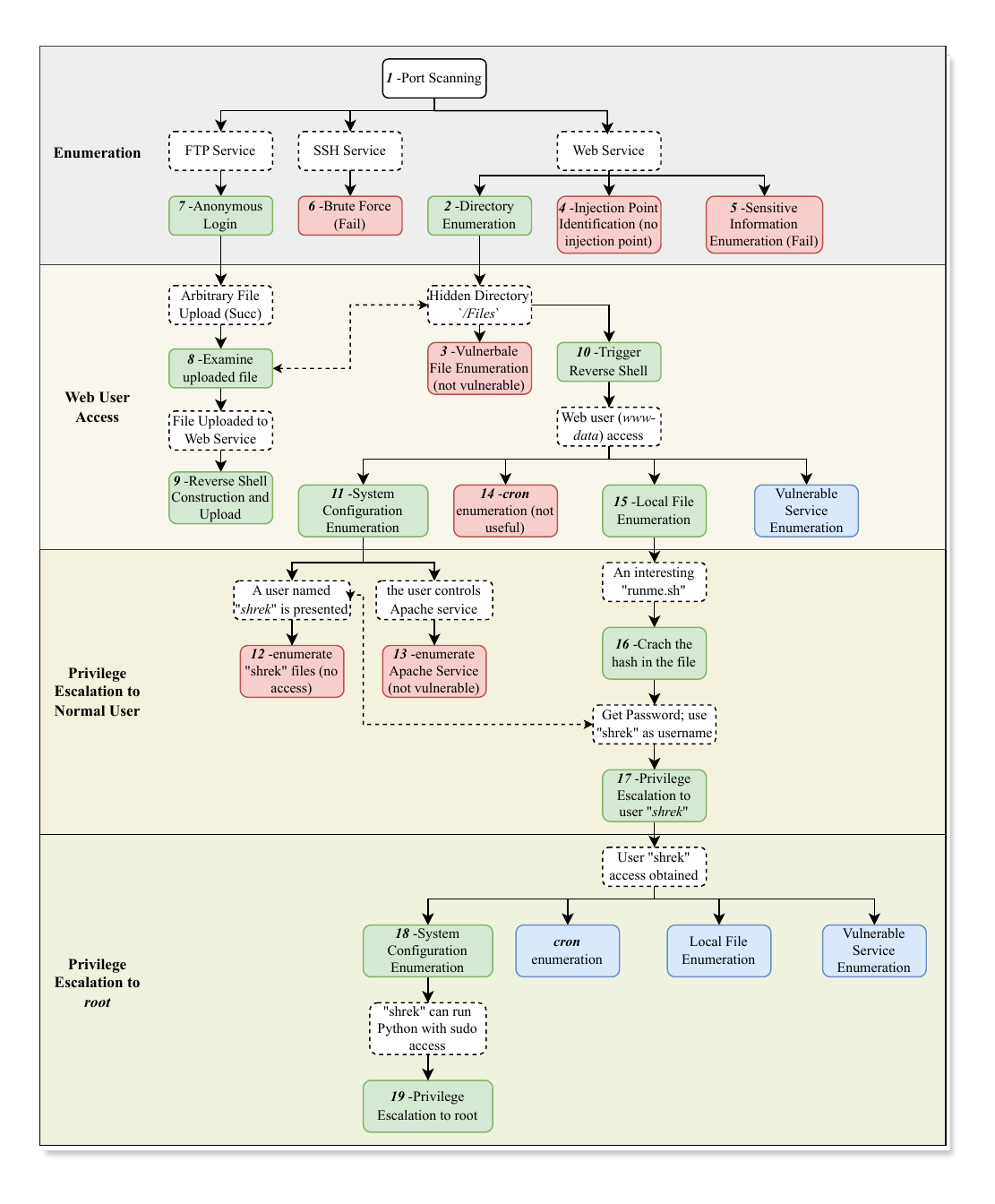}
 \vspace{-40pt}
 \caption{A complete PTT example on the testing target \textit{Vulnhub-Hackable II}}
	\label{fig:complete-task-tree}
\end{figure*}

\edit{To demonstrate the PTT Generation Process in its entirety, we deploy \tool{} on the benchmark system \textit{Hackable II}. Figure~\ref{fig:complete-task-tree} illustrates the complete PTT. In the figure, solid boxes depict the penetration testing operations generated by \tool{}, whereas dotted boxes outline the findings derived from these operations. Red boxes indicate operations that do not yield significant findings, green boxes denote operations that lead to useful findings, and blue boxes represent operations generated by \tool{} but not executed due to lower priority. For clearer presentation, we label the operations with numbers based on the operation sequences as prioritized by \tool{}.}

\edit{As depicted in Figure~\ref{fig:complete-task-tree}, \tool{} emulates the strategic approach typically employed by human penetration testers, encompassing four steps including enumeration, web user access via reverse shell, and privilege escalation to both normal user and root levels on this particular benchmark machine. Notably, \tool{} demonstrates human-like reasoning by linking findings across different stages. During the \textit{Web User Access} phase, it connects a vulnerability in the FTP service with earlier findings from the web service to facilitate an attack by uploading and triggering a reverse shell via FTP. Similarly, in the \textit{Privilege Escalation to Normal User} phase, \tool{} identifies a user named "shrek" on the system, which it then exploits to crack the password and escalate privileges. These instances illustrate \tool{}'s capability to integrate and leverage disparate pieces of information, mirroring the cognitive processes of human testers. }

\end{document}